\documentclass[aps,prb,twocolumn,groupedaddress,amsmath,amssymb,showpacs, superscriptaddress]{revtex4}
\usepackage{amssymb}
\usepackage{amsmath}
\usepackage{graphicx}
\usepackage{epsfig}
\usepackage{longtable}
\usepackage{epstopdf}
\usepackage{dcolumn}
\usepackage{bm}
\usepackage{cancel}
\begin{document}

\preprint{Contact v.4}
\title{Intrinsic Spin Lifetime of Conduction Electrons in Germanium}
\author{Pengke Li} \email{pengke@ece.rochester.edu}
\affiliation{Department of Electrical and Computer Engineering, University of Rochester, Rochester, NY, 14627}
\author{Yang Song}
\affiliation{Department of Physics and Astronomy, University of Rochester, Rochester, NY, 14627}
\author{Hanan Dery}
\affiliation{Department of Electrical and Computer Engineering, University of Rochester, Rochester, NY, 14627}
\affiliation{Department of Physics and Astronomy, University of Rochester, Rochester, NY, 14627}

\begin{abstract}
We investigate the intrinsic spin relaxation of conduction electrons in germanium due to electron-phonon scattering. We derive intravalley and intervalley spin-flip matrix elements for a general spin orientation and quantify the resulting anisotropy in spin relaxation. The form of the intravalley spin-flip matrix element is derived from the eigenstates of a compact spin-dependent $\mathbf{k}$$\cdot$$\mathbf{p}$ Hamiltonian in the vicinity of the $L$ point (where thermal electrons are populated in Ge). Spin lifetimes from analytical integrations of the intravalley and intervalley matrix elements show excellent agreement with independent results from elaborate numerical methods.
\end{abstract}
\maketitle
\section{Introduction}

Group IV semiconductors are natural material choices for quantum and classical spintronic devices.\cite{Kane_Nature98,Zutic_RMP04,Zutic_PRL06,Dery_Nature07}  Hyperfine interactions are suppressed due to the natural abundance of zero-spin nuclear isotopes. As a result, localized electrons have exceedingly long coherence times at low temperatures.\cite{Jelezko_PRL04,Fodor_JPCM06,Tyryshkin_NatureMater12}  As for conduction electrons, space inversion symmetry precludes their spin relaxation by the Dyakonov-Perel mechanism.\cite{DP} The intrinsic spin lifetime is therefore relatively long, reaching $\sim$10~ns at room-temperature in non-degenerate $n$-type silicon .\cite{Cheng_PRL10,Li_PRL11,Dery_APL11,Tang_PRB12,Song_arXiv12} Combined with the fact that silicon is the material of choice in the semiconductor industry, there is a wide interest in related spin injection experiments.\cite{Appelbaum_Nature07,Jonker_NaturePhysics07,Dash_Nature09,Sasaki_APE11,Ando_APL11}

The motivation for studying spin injection in Ge is similar to Si due to their shared properties and the compatibility with Si-based CMOS technology. Electrical spin injection and extraction in Ge have been recently investigated in lateral spin-transport devices with various doping profiles using nonlocal\cite{Zhou_PRB11} and local\cite{Jain_APL11,Saito_SSC11,Jeon_APL11,Jeon_PRB11,Hanbicki_SSC11,Kasahara_JAP12} Hanle measurements, as well as in heterostructure and nanostructure devices.\cite{Shen_APL10,Liu_NanoLett10} Similar to direct band-gap semiconductors, optical orientation is an additional viable tool to investigate spin properties of electrons and holes in Ge.\cite{Loren_APL09,Rioux_PRB10,Bottegoni_APL11,Guite_PRL11,Hautmann_PRB11,Loren_PRB11,Pezzoli_PRL12} Unlike silicon, optical orientation in Ge is efficient because of the energy proximity between the direct and indirect gaps. Spin-polarized electrons are first photoexcited to the $\Gamma$~valley and then they relax via ultrafast spin conserving scattering to the conduction band edges in one of the four $L$~valleys (located $\sim$140~meV below the zone center $\Gamma$-valley).\cite{Pezzoli_PRL12}


Theoretical efforts in the early days\cite{Hasegawa_PR60, Ruth_PR60} were motivated by low-temperature electron spin resonance experiments that studied the $g$
factor and spin-lattice relaxation of localized electrons in donor states.\cite{Feher_PRL_59, Wilson_PR64, Gershenzon_PSS70} On the other hand, little attention was paid to conduction electrons whose spin relaxation is mediated by the Elliott-Yafet mechanism.\cite{Elliott_PR54,Yafet_1963} By analyzing the space inversion and time reversal symmetries of the $L$ point, Yafet deduced a $T^{7/2}$ temperature dependence of the spin relaxation rate due to intravalley electron scattering with acoustic phonons.\cite{Yafet_1963} Kalashnikov extended Yafet's theory to various statistical distributions and scattering mechanisms.\cite{Kalashnikov_67} Chazalviel investigated spin flips due to electron-impurity scattering using effective spin-orbit coupling parameters that resemble the treatment in III-V  semiconductors.\cite{Chaza_JPCS75} Most recently, Tang \textit{et al.} have used a tight-binding model to calculate the intrinsic spin relaxation of conduction electrons in strained Ge.\cite{Tang_PRB12}

In this paper, we present a theory of spin-flip processes due to electron-phonon scattering in Ge. Two distinctive contributions are present in this work. First, we find the spin orientation dependence of spin-flip matrix elements. This dependence leads to anisotropy in spin relaxation and it is instrumental in analyzing measurements where the orientation of injected spins is set by the shape and magnetocrystalline anisotropy of ferromagnetic contacts or by the propagation and helicity of a circularly-polarized light beam. An interesting result of the analysis is that most of the intrinsic spin relaxation of conduction electrons in Ge can be explained by coupling of the lowest conduction band to the upper conduction bands (rather than to the upper valence bands  which is the typical case in most semiconductors). The second contribution of this work is the derivation of a spin-dependent $\mathbf{k}$$\cdot$$\mathbf{p}$ Hamiltonian in the vicinity of the $L$ point (conduction band edge). This compact Hamiltonian model exquisitely captures the signature of spin-orbit interaction on electronic states and it can be extended to study confined Ge structures using an expanded basis of envelope functions.\cite{Gershoni_JQE93}

This paper is organized as follows. Section~\ref{sec:general} discusses the symmetry of wavefunctions and their effects on spin-flip processes. Section \ref{sec:III} provides a quantitative discussion of intervalley spin flips due to scattering with zone-edge phonons. Using group theory, we derive the spin-flip matrix elements and calculate the resulting spin lifetime. A comparison with numerical calculations is also provided. In Sec. \ref{sec:IV} the method of invariants\cite{Bir_Pikus_Book, Ivchenko_Pikus_Book,Winkler_Book,Winkler_PRB10} is used to construct a spin-dependent $\mathbf{k}$$\cdot$$\mathbf{p}$ Hamiltonian around the $L$~point. We solve this Hamiltonian to get the energy bands and spin-dependent wavefunctions. This information is then used in Sec. \ref{sec:V} to study intravalley spin flips due to scattering with long-wavelength acoustic phonons. Section \ref{sec:VI} summarizes our findings.

\section{Symmetry Considerations}\label{sec:general}

\begin{figure}
\includegraphics[width=9cm] {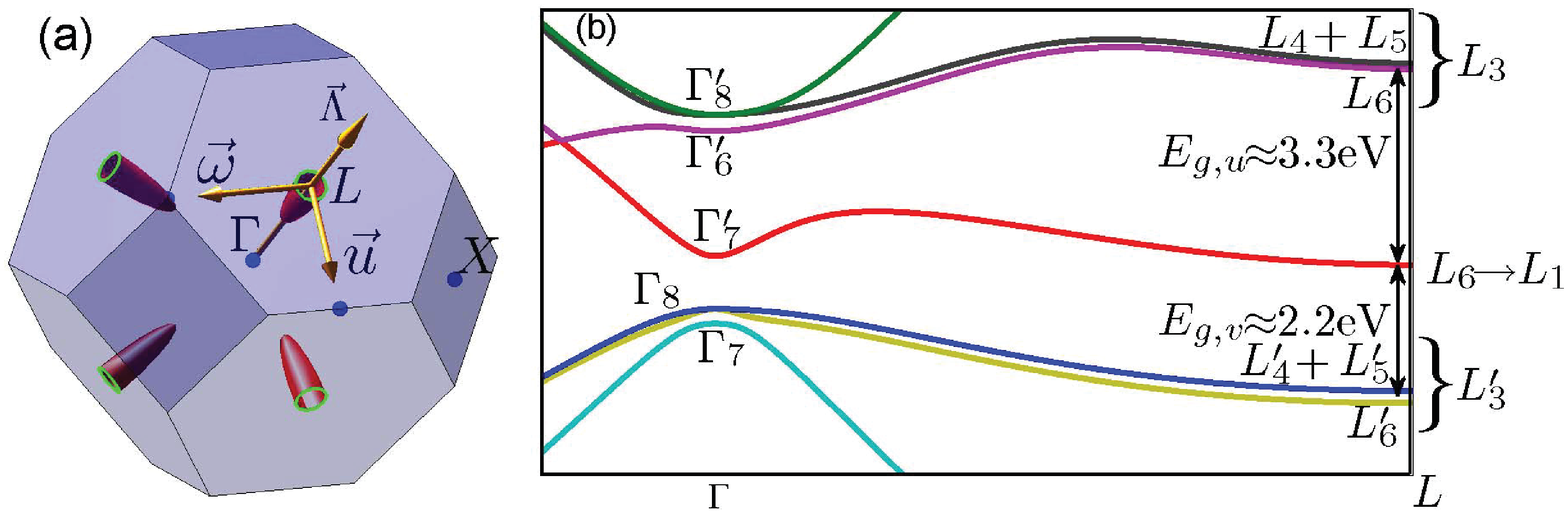}
\caption{ (a)~ The four conduction-band valleys of Ge. These valleys have ellipsoidal energy surfaces and their centers are located at the zone-edge $L$~points. (b)~ Band structure of Ge along high-symmetry crystallographic axes. The conduction-band edge is indicated by the irreducible representation $L_1$ ($L_6$) in single (double) group notation.\cite{Cardona_Book,Dresselhauses_Jorio_Book}
\label{fig:Band}}
\end{figure}

Spin flips during an electron scattering are described by the matrix element $\langle{\mathbf{k},\Uparrow}|H_i|{\mathbf{k}',\Downarrow}\rangle$ where $H_i$ denotes the scattering interaction and $\mathbf{s}=\{\Uparrow,\Downarrow\}$ are the spin-up and spin-down states. In a multivalley conduction band this scattering represents an \textit{intravalley} or an \textit{intervalley} process depending on whether the electron wavevectors $\mathbf{k}$ and $\mathbf{k}'$ reside in the same or different valley(s), respectively.  Figure~\ref{fig:Band}(a) shows the four valleys in the conduction band edge of Ge. If the crystal possesses an inversion center then each band at wavevector $\mathbf{k}$ is spin degenerate and we can define its states with respect to the spin orientation $\mathbf{n}$,
\begin{eqnarray}
 &\langle \mathbf{k}, \mu, \Uparrow |  \bm{\sigma}\cdot \hat{\mathbf{n}} | \,\mathbf{k}, \mu, \Uparrow \rangle  \equiv -\langle \mathbf{k}, \mu, \Downarrow |  \bm{\sigma}\cdot \hat{\mathbf{n}} |\, \mathbf{k}, \mu, \Downarrow \rangle \geq 0&\,,\,\nonumber  \\
 & \langle \mathbf{k}, \mu, \Uparrow |  \bm{\sigma}\cdot \hat{\mathbf{n}} | \,\mathbf{k}, \mu, \Downarrow \rangle \equiv 0\, ,&\label{eq:spin_up_down_states}
\end{eqnarray}
where $\mu$ is the band index. In the following we omit this index and deal only with the lowest conduction band. Mixed by spin-orbit interaction, the spin-up and spin-down states ($\Uparrow, \Downarrow$) in Eq.~(\ref{eq:spin_up_down_states}) are not pure spin states ($\uparrow, \downarrow$). They are written as\cite{Yafet_1963}
\begin{eqnarray}
|{\mathbf{k},\Uparrow}\rangle &=& e^{i\mathbf{k}\cdot\mathbf{r}}\sum_{\ell} (a_{\mathbf{k}}^{\ell}|{\ell}_\uparrow\rangle+b_{\mathbf{k}}^{\ell}|{\ell}_\downarrow\rangle),
\label{eq:spin_state_up}\\
|{\mathbf{k},\Downarrow}\rangle &=& e^{i\mathbf{k}\cdot\mathbf{r}}\sum_{\ell} ( {a_{\mathbf{k}}^{\ell}}^*|{\ell}_{\downarrow} \rangle - {b_{\mathbf{k}}^{\ell}}^*|{\ell}_\uparrow\rangle),
\label{eq:spin_state_down}
\end{eqnarray}
where $\ell$ runs over a basis of periodic Bloch functions. Using the crystal and time reversal symmetries at the four $L$-valley centers, we can identify a handful of periodic functions that describe the spin properties of conduction electrons in Ge. The analysis relies on their transformation properties under symmetry operations.

Representations of conduction (valence) states in the $L$~point have even (odd) parity under space inversion operation. In the notation of single group theory, there are 6 irreducible representations (IR) of the $L$~point space group.\cite{Cardona_Book,Dresselhauses_Jorio_Book} The lowest conduction band is non-degenerate and belongs to $L_1$ (a one-dimensional IR). Figure~\ref{fig:Band}(b) shows that nearby bands are pairs of valence and conduction bands. They are represented, respectively, by the two-dimensional IRs $L_3'$ and $L_3$. If the crystal potential is vanishingly small then the $L$~point energies of these five bands ($L_1$, $L_3$ and $L_3'$) are degenerate. However, the crystal potential in Ge splits these bands into three sets and the relatively large energy separation from $L_1$ will be shown to result in a very slow intravalley spin relaxation process. In comparison, the six conduction band valleys in Si are located close (in energy and wavevector) to the two-band degeneracy at the $X$~point. This degeneracy leads to a spin hot-spot along certain directions in the square boundary faces of the Brillouin zone,\cite{Cheng_PRL10,Li_PRL11,Song_arXiv12} and to a unique behavior of intravalley spin relaxation in Si.\cite{Song_arXiv12} This distinct difference between Si and Ge merits independent treatments of the spin relaxation.

The wavefunctions of conduction electrons, $|{\mathbf{k},\Uparrow (\Downarrow)}\rangle$,  include small contributions from states of remote bands. In the $\mathbf{k}$$\mathbf{\cdot}$$\mathbf{p}$  theory there are two first-order terms related to signatures of the spin-orbit interaction,
\begin{eqnarray}
\label{eq:perturbation_ki}
&&H_{SO} = \frac{\hbar}{4m_0^2c^2}\bm{\nabla} V\times \mathbf{p}\cdot \hat{\sigma}, \\
\label{eq:perturbation_k}
&&H_{SO}^k = \frac{\hbar^2}{4m_0^2c^2}\bm{\nabla} V\times \mathbf{k}\cdot \hat{\sigma}.
\end{eqnarray}
$H_{SO}^k$ transforms as a polar vector ($\bm{\nabla}V$) and can couple between odd and even states. In our case, the coupling is between states of $L_1$ and $L_3'$ symmetries that represent, respectively, the lowest conduction band and upper valence bands. On the other hand, $H_{SO}$ transforms as an axial vector ($\bm{\nabla} V\times \mathbf{p}$) and can couple the even states of $L_1$ and $L_3$ (lowest and upper conduction bands). The dimensionality of $L_3'$ or $L_3$ (2D IRs) is such that each is coupled to $L_1$ by two components of a vector that lie perpendicular to the valley axis [i.e., parallel to the hexagonal boundary faces at the zone edge; see Fig.~\ref{fig:Band}(a)].

The aforementioned $k$-independent coupling between $L_1$ and $L_3$ zone-edge states can be used to find new eigenstates that diagonalize the $L$ point Hamiltonian ($\mathbf{k}$=0) in the presence of spin-orbit coupling,
\begin{eqnarray}
\label{eq:L6}
|L_1 \rangle &\rightarrow& | L_1,s \rangle + \sum_{s'=\uparrow,\downarrow} \sum_{\ell} \frac{|L_{\ell},s'\rangle    \langle L_{\ell},s' | H_{SO} | L_1,s \rangle}{E_{\ell} - E_{L_1}} \nonumber \\ &\approx& |L_1,s\rangle + b_1 | L_3,s \rangle + b_2 | L_3,-s \rangle
\end{eqnarray}
where $b_{1,2} \ll 1$ are $k$-independent parameters, $-s$ is the opposite spin direction of $s$, and $|L_{\ell},s\rangle$ denotes $|L_{\ell} \rangle \otimes |s\rangle$. Slightly away from the valley center, the signature of $L_3$ on conduction electronic states outweighs the signature of $L_3'$ via $H_{SO}^k$ (since $k \ll 2\pi/a$ where $\mathbf{k}$ is measured from the $L$ point and $a$ is the lattice constant).  This property implies stronger spin-orbit coupling with the upper conduction bands than with the upper valence bands. Therefore, we start with a study of the intervalley spin relaxation where the effect of $H_{SO}$ enters at the lowest order (the spin-flip amplitude does not depend on relative value of the wavevectors with respect to their valley centers).

\section{Intervalley Spin-Flip Transitions}
\label{sec:III}
The intervalley scattering in Ge is governed by electron interaction with zone-edge phonons close to the $X$~point. For example, the $L_{111}$ and $L_{11\overline{1}}$ valley centers [$\mathbf{k}=\pi/a(1,1,\pm 1)$] are connected by the wavevector $X_{001}$ [$\mathbf{k}=2\pi/a(0,0,1)$]. These zone-edge phonons belong to three 2D IRs: $X_3$ (TA, 10 meV), $X_1$ (LA \& LO, 29 meV) and $X_4$ (TO,~33~meV),\cite{Nilsson_PRB72} with their polarizations and energies written in the parentheses. If the transition involves valleys that we denote by $L$ and $L_t$, then the selection rule for electron scattering reads $L_1\otimes L_{1t} = X_1\oplus \bcancel{X_3}$ where $X_3$ is forbidden by time reversal symmetry.\cite{Lax_61,Streitwolf_70}

Including the spin degree of freedom, it is convenient to use double group theory where $L_{1}$ is replaced by $L_{6}$. The symmetry properties of the renormalized state in Eq.~(\ref{eq:L6}) are described by the $L_6$ IR. The new selection rule reads\cite{Tang_PRB12}
\begin{eqnarray}
\label{eq:phonons}
L_6 \otimes L_{6(t)} = 2X_1 \oplus X_4 \oplus \bcancel{X_3} ,
\end{eqnarray}
and it means that three independent scattering parameters are needed to fully describe intervalley scattering (two are related to the $X_1$ symmetry and one to $X_4$). Scattering with zone-edge phonons of $X_3$ symmetry is excluded at the lowest order by time reversal symmetry. This selection rule does not provide information on the spin orientation dependence. To overcome this shortcoming, we can work with IR matrices rather than their traces.\cite{Song_arXiv12} Technical details of applying this approach in Ge are given in Appendix A and below we present the main findings.

Spin flips due to scattering with phonons of $X_4$ symmetry are described by a single independent nonvanishing matrix-element constant which we denote by $D_{X_4,s}$. On the other hand, of the two independent nonvanishing matrix elements of $X_1$, one is attributed to spin-flip transitions ($D_{X_1,s}$), and the other to spin-conserving scattering ($D_{X_1,m}$). Most important in analyzing experiments is the dependence of the scattering amplitude on the spin orientation of electrons. The spin orientation, $\hat{\mathbf{n}}$, is described by a polar angle from the $+z$ direction ($\theta$) and an azimuthal angle in the $xy$ plane measured from the $+x$ direction ($\phi$). For a spin-flip transition between $L_{111}$ and $L_{11\overline{1}}$ valleys (via a zone-edge phonon with wavevector $X_{001}$), the square amplitude of the matrix element reads
\begin{eqnarray}
\label{eq:Ms}
&&\sum_{j=1,2} |\langle \mathbf{k}_{Lt}, \Downarrow_{\mathbf{n}}\! |{H}_{X^j_i}| \mathbf{k}_L, \Uparrow_{\mathbf{n}} \rangle|^2 \\
=&&\left\{\begin{array}{ll}2D_{X_1,s}^2(1+\cos^2\theta + \sin 2\phi \sin^2\theta ) &\mbox{ if }
i = 1 \\ 2D_{X_4,s}^2 \sin^2\theta  & \mbox{ if } i=4\end{array}\right. \,.\nonumber
\end{eqnarray}
$j$ sums the two degenerate branches of each $X$~point IR. For the remaining five transitions between other combinations of $L$~points, Eq.~(\ref{eq:Ms}) varies according to a straightforward coordinate transformation. These results are summarized in Table~\ref{tab:M_SL} (also see discussion of Eq.~(\ref{eq:Ms_xyz}) in Appendix~A). As seen by the angular dependence in Eq.~(\ref{eq:Ms}), spin-flip transitions are forbidden for specific choices of the spin orientations. For example, if the spin is oriented parallel to the phonon wavevector ($z$-axis in this case and $\theta = 0$), then the spin-flip amplitude vanishes for the $X_4$ symmetry. Below, we will quantify the effect of spin orientation on the spin lifetime.

\begin{table} 
\renewcommand{\arraystretch}{2}
\tabcolsep=0.1cm
\caption{\label{tab:M_SL}
$\sum_{j=a,b}|M_{X^{j}_i} (\mathbf{k}_L,\Uparrow_\mathbf{n}; \mathbf{k}_{Lt},\Downarrow_\mathbf{n})/ 2D_{X_i}|^2$ for intervalley spin-flip transitions of all the six valley-to-valley configurations.
}
\begin{tabular}{ccc}
\hline \hline
$L\leftrightarrow L_t$ & $X_1$ & $X_4$ \\ \hline
$L_{111}\leftrightarrow L_{11\overline{1}}$ & $1+\cos^2\theta+\sin^2\theta\sin2\phi$ & $1-\cos^2\theta$  \\
$L_{111}\leftrightarrow L_{1\overline{1}1}$  & $1+\sin^2\theta\sin^2\phi+\sin2\theta\cos\phi$ & $1-\sin^2\theta\sin^2\phi$  \\
$L_{111}\leftrightarrow L_{\overline{1}11}$  & $1+\sin^2\theta\cos^2\phi+\sin2\theta\sin\phi$ & $1-\sin^2\theta\cos^2\phi$  \\
$L_{\overline{1}11}\leftrightarrow L_{11\overline{1}}$  &$1+\sin^2\theta\sin^2\phi-\sin2\theta\cos\phi$ & $1-\sin^2\theta\sin^2\phi$  \\
$L_{\overline{1}11}\leftrightarrow L_{1\overline{1}1}$  & $1+\cos^2\theta-\sin^2\theta\sin2\phi$ & $1-\cos^2\theta$  \\
$L_{1\overline{1}1}\leftrightarrow L_{11\overline{1}}$  & $1+\sin^2\theta\cos^2\phi-\sin2\theta\sin\phi$ & $1-\sin^2\theta\cos^2\phi$  \\
\hline \hline
\end{tabular}
\end{table}

Having the spin-flip matrix elements, one can readily calculate the intervalley spin relaxation rate by integration over the Brillouin zone,
\begin{eqnarray}
\frac{1}{\tau_{s,\nu}}=\frac{2\pi\hbar}{\varrho N_c} \int d^3 \mathbf{k} \left.\frac{\partial f(E)}{\partial E}\right|_{E_\mathbf{k}} \int  \frac{d^3\mathbf{k}'}{(2\pi)^3} \frac{\big|  M_\nu^{sf}(\mathbf{k},\mathbf{k}') \big|^2}{\Omega_{\nu}(\mathbf{q})}
\quad&&\label{eq:ts_phonon}\\
\sum_{\pm}(n_{\nu,\mathbf{q}} + \tfrac{1}{2} \pm \tfrac{1}{2}) \delta( E_{\mathbf{k}'}  - E_{\mathbf{k}} \pm \Omega_{\nu}(\mathbf{q}) ).&&\nonumber
\end{eqnarray}
$\varrho=5.323$~g/cm$^3$ is the crystal density, $f(E)$ is the statistical distribution of electronic states, and $N_c=~\!\!\!\int d^3 \mathbf{k} \left.\partial f(E)/\partial E\right|_{E_\mathbf{k}}$ is an effective density of states constant. $\nu$, $\mathbf{q}=\mathbf{k}-\mathbf{k}'$, $\Omega_{\nu}(\mathbf{q})$ and $n_{\nu,\mathbf{q}}$ denote, respectively, the phonon mode, its wavevector, energy and Bose-Einstein distribution. The $\pm$ refers to phonon emission and absorption processes. $M_\nu^{sf}(\mathbf{k},\mathbf{k}')=~\!\!\langle \mathbf{k}',\Downarrow\!\!| M_{\nu,\mathbf{q}} |\mathbf{k},\Uparrow\rangle$  is the spin-flip matrix element.

Using the spin-flip matrix elements in Table~\ref{tab:M_SL}, we calculate the intervalley spin relaxation rate for a Boltzmann distribution of electrons,
\begin{eqnarray}
\!\!\!\frac{1}{\tau_{s,inter}} =\frac{4}{3} \left(\!\frac{2m_d}{\pi}\!\right)^{\!\!\!\frac{3}{2}}\!\!\sum_{i=1,4}\frac{ A_i(\theta,\phi) D_{X_{i},s}^2}{\hbar^2 \varrho \sqrt{\Omega_i}} \frac{\vartheta(y_i)}{\exp(y_i)\!-\!1}.\,\,\,
\label{eq:ts_inter}
\end{eqnarray}
$m_d$=$\sqrt[3]{m_l m_t^2}$ is the effective electron mass where $m_t$$\,$$\approx$$\,$0.08$m_0$ and $m_l$$\,$$\approx$$\,$1.6$m_0$ are, respectively, the transverse and longitudinal components. $\vartheta$($y_i$=$\Omega_i$/$k_BT$)$\!\!$ = $\!\!\sqrt{y_i}$$\exp(y_i/2)$$K_{-1}(y_i/2)$ is associated with the modified Bessel function of the second kind. This term slightly depends on temperature and varies between $2$ and $4$ in the temperature range between 10~K and 400~K (for both phonon energies). On the other hand, most of the temperature dependence of the intervalley relaxation rate comes from the thermal population of zone-edge phonons (exponent term in the denominator). This population is strongly suppressed at low temperatures. Finally, the scattering constants are  $D_{X_{1},s} = 35\,\text{meV}/\text{\AA}$ and $D_{X_{4},s} = 46\,\text{meV}/\text{\AA}$, quantified by numerical results of empirical pseudopotential method,\cite{Chelikowsky_PRB76} adiabatic bond-charge model,\cite{Weber_PRB77} and rigid-ion approximation.\cite{Allen_PRB81} They are respectively weighted by $A_1(\theta,\phi)$ and $A_4(\theta,\phi)$ that include the dependence on the spin orientation. We discuss their values for several general cases.

\textit{no-strain or $[100]$-strain}: The four $L$-valleys in the lowest conduction band are degenerate and transitions between all six pairs of valleys are equivalent. The anisotropy in spin relaxation due to intervalley scattering between two valleys is compensated by opposite anisotropy of other pairs. The sum of expressions in each of the two columns of Table~\ref{tab:M_SL} is independent of $\theta$ and $\phi$,
\begin{eqnarray}
A_1 = 8\,, \qquad \qquad A_4=4\,. \label{eq:A_bulk_and_001}
\end{eqnarray}
As shown next, when the symmetry between different valleys is broken, the dependence of the intervalley matrix elements on spin orientation lends itself to a measurable anisotropy in the spin lifetime.

\textit{$[111]$~strain}: The case of uniaxial compressive strain results in a single low-energy valley (along the strain axis) and three higher energy valleys.  At relatively large strain levels ($\sim$1\%), the energy split is large enough to quench the intervalley spin relaxation mechanism.\cite{Tang_PRB12}  This effect amounts to assigning $1/\tau_{s,inter}=0$. On the other hand, in biaxial compressive strain configuration (or uniaxial tensile strain) three of the valleys shift down in energy and one valley shifts up. Excluding transitions with the $L_{111}$ valley (considering the last three lines in Table~\ref{tab:M_SL}) we get
\begin{eqnarray}
A_1\!=\!\frac{16\!-\!4\sin^{\!2}\!\theta \sin\!2\phi \!-\! 4\sin\!2\theta(\!\cos\!\phi\!+\!\sin\!\phi\!)}{3},\,A_4\!=\!\frac{8}{3}.\,\,\, \label{eq:A_111}
\end{eqnarray}
This strain configuration restores the anisotropy in spin relaxation due to electron scattering with $X_1$ zone-edge phonons. By changing the spin orientation from the strain axis to its perpendicular plane, $\tau_{s,inter}$ drops by $\sim$50\% [changing $A_1$ from 8/3 to 20/3 in Eq.~(\ref{eq:ts_inter})] .

\textit{$[110]$-strain}: This strain configuration is optimal for detection of the anisotropy since intervalley transitions are effective from a single pair of valleys. Consider, for example, the case that $L_{111}$ and $L_{11\overline{1}}$ valleys shift sufficiently down in their energy. Then, only the first term in Table~\ref{tab:M_SL} represents the intervalley scattering and we get
\begin{eqnarray}
A_1 = 2(1+\cos^2\theta + \sin 2\phi \sin^2\theta)\,,\,\,\,\,A_4 = 2\sin^2\theta\,.\,\,\,\, \label{eq:A_110}
\end{eqnarray}
The anisotropy in spin relaxation is now caused by electron scattering with both types of zone-edge phonons. By changing the spin orientation from the strain axis to its perpendicular plane, $\tau_{s,inter}$ [Eq.~(\ref{eq:ts_inter})] is doubled.

Other than strain, it should also be possible to observe the anisotropy by applying electric fields of few kV/cm along the mentioned directions. Here, valley repopulation will result in preferential scattering from hot-to-cold valleys.\cite{Li_PRL12} Finally, by averaging over spin orientations in Eqs.~(\ref{eq:A_111})-(\ref{eq:A_110}), the spin lifetime [Eq.~(\ref{eq:ts_inter})] with two (three) low-energy valleys is about 3 (3/2) times longer than that of the unstrained case. The reason is that electrons can scatter to one (two) valleys rather than three.

\begin{figure}
\includegraphics[width=8.5cm,height=4.5cm] {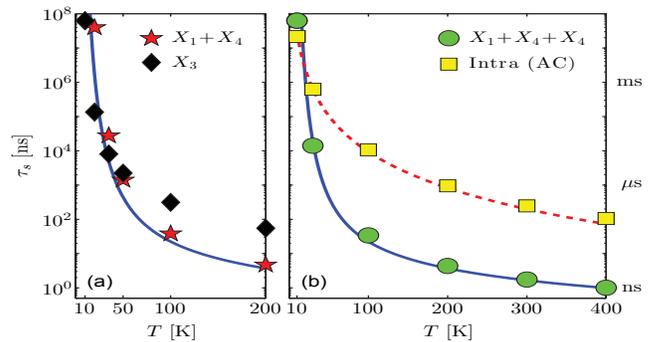}
\caption{Temperature dependence of the intrinsic spin lifetime in unstrained Ge for spins oriented along the $z$-crystallographic axis. The solid lines follow Eq.~(\ref{eq:ts_inter}) and denote the spin lifetime from intervalley scattering with $X_1$ and $X_4$ phonons [the temperature range is extended in (b)]. At room temperature, the spin lifetime of this scattering is $\sim1\,\text{ns}$. The markers in (a) are numerical results. Black diamonds denote contributions from $X_3$ phonons and red pentagrams from $X_1$ and $X_4$ phonons. The green circle markers in (b) show the combined effect. The dash red line in (b) denotes the spin lifetime from intravalley scattering with long-wavelength acoustic phonons [Eq.~(\ref{eq:ts_intra})]. Yellow square markers are the respective numerical results.
\label{fig:ts}}
\end{figure}

The solid blue curve in Fig. \ref{fig:ts}(a) shows the temperature dependence of the intervalley spin lifetime in unstrained bulk Ge [Eq.~(\ref{eq:ts_inter}) with $A_1$=8 and $A_4$=4]. We have also performed numerical integrations of Eq.~(\ref{eq:ts_phonon}) following the procedure in Ref.~[\onlinecite{Cheng_PRL10}]. The numerical calculation of the electron-phonon matrix element follows a rigid-ion approximation.\cite{Allen_PRB81,Li_PRL10} Whereas this numerical approach is not transparent compared with the group theory analysis, it takes into account higher-order corrections (variation of the matrix element when departing from the center of the valley). Nonetheless, the numerical results for scattering with $X_1$ and $X_4$ phonons [red pentagram markers in Fig. \ref{fig:ts}(a)] show that the zeroth-order analytical calculation is sufficient [Eq.~(\ref{eq:ts_inter})]. The black diamond markers denote numerical results due to scattering with $X_3$ phonons. While their zeroth-order contribution vanishes by time reversal symmetry [Eq.~(\ref{eq:phonons})], their first-order contribution (linear in wavevector) becomes important at low temperatures. The reason is that their low energy ($\sim$10~meV) leads to a much larger population at low temperatures compared with that of $X_1$ and $X_4$ phonons (whose energies are $\sim$30~meV).

\section{$L$~point Hamiltonian}
\label{sec:IV}
In this section, we expand the wavefunctions of electrons using the $L$~point basis states. Important signatures of spin-orbit coupling on these wavefunctions will be identified and correlated with intravalley and intervalley spin relaxation processes. Near the valley center, the wavefunction is approximated by
\begin{eqnarray}
|{\mathbf{k},s}\rangle \!=\! \Big[  \sum_{\gamma=1,3,3'} \mathbf{C}_{\gamma}\!(\mathbf{k},\!s) | \mathbf{L}_{\gamma} \rangle  \Big]\! e^{i(\mathbf{k}_L+\mathbf{k})\cdot\mathbf{r}},\,\,\,\,\,\, \label{eq:L_gen}
\end{eqnarray}
where following Eqs.~(\ref{eq:spin_state_up})-(\ref{eq:spin_state_down}), we write
\begin{eqnarray}
\mathbf{C}_{\gamma}\!(\mathbf{k},\!\Uparrow) | \mathbf{L}_{\gamma} \rangle \!\!&=&\!\!\!\! \sum_{m=1}^{N_\gamma} \! a_{m,\gamma}\!(\mathbf{k}) | L_{\gamma}^{m}, \uparrow \rangle + b_{m,\gamma}\!(\mathbf{k}) | L_{\gamma}^{m}, \downarrow \rangle , \nonumber \\
\mathbf{C}_{\gamma}\!(\mathbf{k},\!\Downarrow) | \mathbf{L}_{\gamma} \rangle \!\!&=&\!\!\!\! \sum_{m=1}^{N_\gamma} \! a_{m,\gamma}^{\ast}\!(\mathbf{k})  | L_{\gamma}^{m}, \downarrow \rangle - b_{m,\gamma}^{\ast}\!(\mathbf{k}) | L_{\gamma}^{m}, \uparrow \rangle .\,\,\,\,\,\,\,\,\,
\label{eq:Leigen}
\end{eqnarray}
Totally, we consider 10 spin-dependent basis states: two from the non-degenerate lowest conduction band ($N_1$=1), and four from either the upper conduction or valence bands (each being two-band degenerate in the absence of spin-orbit coupling, $N_3$=$N_{3'}$=2). The coefficients are eigenvectors of the Hamiltonian,
\begin{eqnarray}
\left(\!\!
\begin{array}{ccc}
H_{33}\!+\!E_{g,u}   & H_{13}^{\dagger}     & H_{33'}            \\
H_{13}               & H_{11}               & H_{13'}            \\
H_{33'}^{\dagger} & H_{13'}^{\dagger}       & H_{3'3'}\!-\!E_{g,v}
\end{array}
\!\!\right) \!\!
\left( \!\!\begin{array}{c}
\mathbf{C}_3   \\
\mathbf{C}_1   \\
\mathbf{C}_{3}\!\!\!'\,
\end{array}\!\!\right) \!
\!=\!E\!
\left(\!\!\begin{array}{c}
\mathbf{C}_3   \\
\mathbf{C}_1   \\
\mathbf{C}_{3}\!\!\!'\,
\end{array}\!\!\right)\!\!,\,\,\,
\label{eq:10by10}
\end{eqnarray}
where $H_{ij}$ is a matrix block denoting the spin and wavevector dependent coupling between basis states with $L_{i}$ and $L_{j}$ symmetries. $E_{g,u}$ and $E_{g,v}$ denote, respectively, the $L$~point energy separations of the lowest conduction band from the upper conduction and upper valence bands [see Fig.~\ref{fig:Band}(b)]. Below, we present the Hamiltonian matrix using the basis functions of the $L_{111}$ point [$\mathbf{k}_L=\pi(1,1,1)/a$]. Matrix forms in the $\langle \bar{1}11\rangle$, $\langle 1\bar{1}1\rangle$, and $\langle 11\bar{1}\rangle$ valleys are derived by trivial coordinate transformation. In addition, to derive a compact matrix form we use a rotated set of Cartesian coordinates,
\begin{eqnarray} \label{eq:cart}
\hat{\mathbf{w}} = \frac{\hat{\mathbf{x}}-\hat{\mathbf{y}}}{\sqrt{2}},\,\hat{\mathbf{u}} = \frac{\hat{\mathbf{x}}+\hat{\mathbf{y}}-2\hat{\mathbf{z}}}{\sqrt{6}},\,\hat{\mathbf{\scriptstyle{\Lambda}}}= \frac{\hat{\mathbf{x}}+\hat{\mathbf{y}}+\hat{\mathbf{z}}}{\sqrt{3}}\,.
\end{eqnarray}
$\hat{\mathbf{u}}$ and $\hat{\mathbf{w}}$ lie parallel to the hexagonal boundary face [Fig.~\ref{fig:Band}(a)]. $\hat{\mathbf{\scriptstyle{\Lambda}}}$ is along the valley axis connecting the $\Gamma$ and $L_{111}$ points.

We construct the Hamiltonian matrix [Eq.~(\ref{eq:10by10})] using the method of invariants.\cite{Bir_Pikus_Book, Ivchenko_Pikus_Book,Winkler_Book,Winkler_PRB10}  Application of this method with relevance to the $L$ point is given in Appendix B. Here we summarize the findings. The lowest conduction band is associated with the identity IR and contributes a trivial 2$\times$2 matrix form ($L_1\otimes L_1=L_1$),
\begin{eqnarray}
H_{11}=\Big[\frac{\hbar^2(k_u^2+k_{w}^2)}{2m_t^{\ast}}  +\frac{\hbar^2k_{\scriptscriptstyle{\Lambda}}^2}{2m_l^{\ast}}\Big] \otimes I_{2\times2}\,.
\label{eq:H_cc}
\end{eqnarray}
$m_t^{\ast}$ and $m_l^{\ast}$ are effective mass parameters representing the effect of remote bands (outside the chosen basis states). Matrix blocks of the upper valence bands or upper conduction bands share a similar form ($L_3\otimes L_3=L_3'\otimes L_3'=L_1+L_2+L_3$),
\begin{eqnarray}
H_{ii}\!=\!\Big[\frac{\hbar^2(k_u^2+k_{w}^2)}{2m_{t,i}^{\ast}} +\frac{\hbar^2k_{\scriptscriptstyle{\Lambda}}^2}{2m_{l,i}^{\ast}}\Big]\!\otimes\!I_{4\times4} +\Delta_{i}\rho_y\!\otimes\!\sigma_{\scriptscriptstyle{\Lambda}},\,\,\,
\label{eq:H_vv}
\end{eqnarray}
where $i=3$ or $i=3'$. The mass parameters have similar meaning as in $H_{11}$. $\Delta_{i}$ denotes the internal spin-orbit coupling between the two $L_{i}$ basis functions.\cite{Tauc_PRL60} $\rho_y=\sigma_y$ originates from the two-band degeneracy in the absence of spin-orbit coupling.

The off-diagonal matrix block $H_{13'}$ denotes the coupling between the lowest conduction band and upper valence bands. Its form follows from $L_1\otimes L_3'=L_3'$,
\begin{eqnarray}
H_{13'}&=& P \Big(k_{w}[0,1]-k_u[1,0]\Big)\otimes I_{2\times2} \label{eq:H_cv} \\
&+&i\alpha\Big[(\mathbf{k}\times\mbox{\boldmath$\sigma$})_{w}\otimes [0,1] - (\mathbf{k}\times\mbox{\boldmath$\sigma$})_{u}\otimes [1,0]\Big],  \nonumber
\end{eqnarray}
where $[1,0]$ and $[0,1]$ are ordinary $1\times2$ matrices. Their Kronecker products with ${2\times2}$ matrices indicate that $H_{13'}$ is a $2\times4$ matrix. $P$ and $\alpha$  are two independent matrix element constants that originate from the $\mathbf{k}\cdot\mathbf{p}$ and $H_{SO}^{\mathbf{k}}$ perturbation terms, respectively. The coupling matrix of the lowest and upper conduction bands is wavevector independent and it follows from $L_1\otimes L_3=L_3$,
\begin{eqnarray}
\label{eq:H_cu}
H_{13}&=&i\Delta_{L}\Big(\sigma_u\otimes [1,0]+\sigma_w\otimes [0,1]\Big),
\end{eqnarray}
where $\Delta_{L}$ denotes the direct spin-orbit coupling between these bands. Finally, the 4$\times$4 coupling matrix between the upper valence and conduction bands follows from $L_3'\otimes L_3=L_1'+L_2'+L_3'$,
\begin{eqnarray}
H_{33'}&=& \left[ P_1(  ik_u\rho_y - k_w I_{2\times2} ) + P_2k_{\scriptscriptstyle{\Lambda}}\rho_x\right]\otimes I_{2\times2},\,\,\,\,
\label{eq:H_uv}
\end{eqnarray}
where we have neglected the $H_{SO}^{\mathbf{k}}$ coupling between these bands since it plays a negligible role in the spin relaxation of conduction-valley electrons. Table~\ref{tab:params} in Appendix B lists the values of all parameters in Eqs.~(\ref{eq:H_cc})-(\ref{eq:H_uv}).

Given the relatively large $L$~point energy gaps, the energy dispersion of electrons in $L$-valleys is well approximated by eigenvalues of $H_{11} + H_{13'}H_{13'}^{\dagger}/E_{g,v} - H_{13}H_{13}^{\dagger}/E_{g,u}$,
\begin{eqnarray}
E_c= \frac{2\Delta_L^2}{E_{g,u}} + \frac{\hbar^2(k_u^2+k_w^2)}{2m_{t}}+\frac{\hbar^2k_{\scriptscriptstyle{\Lambda}}^2}{2m_{l}}\,.
\label{eq:eigen_energy_c_111_transformed}
\end{eqnarray}
The constant energy shift is due to the direct spin-orbit coupling with the upper conduction bands. The effective mass parameters are
\begin{eqnarray}
\frac{1}{m_t} = \frac{1}{m_t^{\ast}}+\frac{2P^2+2\alpha^2}{\hbar^2E_{g,v}}\,, \qquad  \frac{1}{m_l} = \frac{1}{m_l^{\ast}}+\frac{4\alpha^2}{\hbar^2E_{g,v}}\,. \nonumber
\label{eq:eigen_energy_c_111_transformed}
\end{eqnarray}
About half of the anisotropy between the transverse and longitudinal effective masses in Ge ($m_t$$\,$$\approx$$\,$0.08$m_0$ and $m_l$$\,$$\approx$$\,$1.6$m_0$) is set by the spin independent coupling with the upper valence bands ($P=9\,\text{eV}\cdot\text{\AA}$). The spin-orbit coupling signatures on the energy dispersion are negligible and can be ignored ($\alpha$=40~meV$\cdot$$\text{\AA}$ and $\Delta_L$=27~meV). On the other hand, the minute effect of spin-orbit coupling on the eigenvectors of Eq.~(\ref{eq:10by10}) sets the timescale for spin relaxation.  Choosing the spin quantization along the valley axis, the spin-up eigenvector along this direction [$\hat{\mathbf{n}}$$\,$=$\,$$\hat{\scriptstyle{\Lambda}}$ in Eq.~(\ref{eq:spin_up_down_states})] reads
\begin{eqnarray}
\mathbf{C}_{1}(\mathbf{k},\Uparrow_{\scriptscriptstyle{\Lambda}}) \!\!\!&=& \!\!\! \Big[ \huge{1},\, g(\mathbf{k}) \Big] + \mathcal{O}(k^2),\,\,\,\, \nonumber \\
\mathbf{C}_{3}(\mathbf{k},\Uparrow_{\scriptscriptstyle{\Lambda}})  \!\!\!&=& \!\!\! \frac{\Delta_L}{E_{g,u}}\!\Big[ 0 ,\,  -1,\,  0,\, i \Big] + \mathcal{O}(k^2) \,,\,\,\,\,\,\, \nonumber \\
\mathbf{C}_{3}\!\!'\,(\mathbf{k},\Uparrow_{\scriptscriptstyle{\Lambda}}) \!\!\!&=& \!\!\! \frac{P}{E_{g,v}}\!\Big[ -k_{u}-i\gamma_3 k_w,\,f_{\!+}(\mathbf{k}), \nonumber \\
\!\!\!&& \!\!\! \qquad \qquad k_w-i\gamma_3 k_u,\, f_{\!-}(\mathbf{k}) \Big]+ \mathcal{O}(k^3).\,\,\,\,\,\,\,\,\,\,\,\, \label{eq:C_values}
\end{eqnarray}
The components of the spin-down eigenvector [$\mathbf{C}_{i}(\mathbf{k},\Downarrow_{\scriptscriptstyle{\Lambda}})$] are readily obtained from space inversion and time reversal relations [Eq.~(\ref{eq:Leigen})]. The $g(\mathbf{k})$ and $f_{\pm}(\mathbf{k})$ functions in $\mathbf{C}_1$ and $\mathbf{C}_{3}\!\!'$ read
\begin{eqnarray}
\!\!\! g(\mathbf{k}) \!\!&=&\!\!  \frac{P^2}{E_{g,v}^2}\big[k_uf_{\!+}(\mathbf{k})-k_w f_{\!-}(\mathbf{k})\big],\,\, \nonumber \\
\!\!\! f_{\!\pm}(\mathbf{k})  \!\!&=&\!  r_{\pm} \left[ \gamma_1 ( k_w -ik_u) \pm i \gamma_2 k_{\scriptscriptstyle{\Lambda}} \right]  ,\,\,\,\,\,\,   \label{eq:gf_func}
\end{eqnarray}
where $r_+$$\,$=$\,$1 and $r_-$$\,$=$\,$$-i$. The $\gamma_j \ll 1$ parameters scale with three of the spin-orbit coupling constants ($\alpha$, $\Delta_{3'}$ and $\Delta_{L}$)
\begin{subequations} \label{eq:gamma_j}
\begin{align}
\gamma_1  &=  \frac{\Delta_L}{E_{g,u}}\frac{P_1}{P} \approx 0.006, \,\,\,\,\,\, \label{eq:Apm} \\
\gamma_2  &= \frac{\alpha}{P} + \frac{\Delta_L}{E_{g,u}}\frac{P_2}{P} \approx  0.005,  \,\,\,\,\,\, \label{eq:Bpm} \\
\gamma_3  &= \frac{\alpha}{P} + \frac{\Delta_{3'}}{E_{g,v}} \approx  0.05,  \,\,\,\,\,\,\,  \label{eq:gammapm}
\end{align}
\end{subequations}
where the internal spin-orbit coupling in the valence band ($\Delta_{3'}$) sets most of the value of $\gamma_3$. Only when the spin is oriented along the valley axis ($\hat{\mathbf{n}}$$\,$=$\,$$\hat{\scriptstyle{\Lambda}}$), this parameter is excluded from the opposite-spin components of $\mathbf{C}_{3}\!\!'\,(\mathbf{k},\Uparrow_{\mathbf{n}})$ [i.e, from the $f_{\!\pm}(\mathbf{k})$ terms in Eq.~(\ref{eq:C_values})]. This behavior will have important consequences on the anisotropy of intravalley spin relaxation. 

\subsection*{Connection between the $L$~point Hamiltonian parameters and spin relaxation}
\label{sec:H_and_ts}
To facilitate a connection between the Hamiltonian eigenvectors and spin relaxation we make use of spin-flip overlap integrals. We show that the direct spin-orbit coupling between the conduction bands ($\Delta_{L}$) plays a key role in setting the intervalley spin relaxation rate (independently derived in Sec.~\ref{sec:III}). On the other hand, we will see that intravalley spin-flip transitions are weaker. To make these connection clear, we write the overlap integral
\begin{eqnarray}\label{eq:overlap1}
\!\!\!&&\!\!\! \mathcal{I}( \mathbf{k}',\Downarrow_{\mathbf{n}}\,;\,\mathbf{k},\Uparrow_{\mathbf{n}}) =  \nonumber \\ && \qquad \sum_{\mu, \gamma}   \big\langle \mathbf{L}_{\mu,\mathbf{k}_{\scriptscriptstyle{L'}}} \,\big|  \mathbf{C}_{\mu}^{\dagger}(\mathbf{k}',\Downarrow_{\mathbf{n}})   \mathbf{C}_{\gamma}(\mathbf{k},\Uparrow_{\mathbf{n}}) \,\big| \mathbf{L}_{\gamma,\mathbf{k}_{\scriptscriptstyle{L}}} \big\rangle,\,\,\,\,\,
\end{eqnarray}
where $\mathbf{k}$ and $\mathbf{k}'$ are measured from the nearby valley center ($\mathbf{k}_{\scriptscriptstyle{L}}$ and $\mathbf{k}_{\scriptscriptstyle{L'}}$). The $bra$ and $ket$ states of this overlap integral include only the periodic Bloch parts in Eq.~(\ref{eq:L_gen}). While the combined phase factor, $\exp{\{i(\mathbf{k}_{\scriptscriptstyle{L}}-\mathbf{k}_{\scriptscriptstyle{L'}}+\mathbf{k}-\mathbf{k}') \cdot \mathbf{r}\}}$, is excluded from the overlap integral, it will be taken into account in the phonon phase when calculating the matrix elements. Using Eq.~(\ref{eq:C_values}), the overlap integrals of electrons in different valleys reads ($\mathbf{k}_{\scriptscriptstyle{L}} \neq \mathbf{k}_{\scriptscriptstyle{L'}}$)
\begin{eqnarray}\label{eq:overlap_inter}
\mathcal{I}_e( \mathbf{k}',\Uparrow_{\mathbf{n}}\,;\,\mathbf{k},\Uparrow_{\mathbf{n}}) \!\!&=&\!\!   c_{1,\mathbf{n}} \nonumber \\
\mathcal{I}_e( \mathbf{k}',\Downarrow_{\mathbf{n}}\,;\,\mathbf{k},\Uparrow_{\mathbf{n}}) \!\!&=&\!\!  c_{3,\mathbf{n}} \frac{\Delta_L}{E_{g,u}} + \mathcal{O}(k^2)\,,\,\,\,\,\,
\end{eqnarray}
where $c_{j,\mathbf{n}}$ are constants of order unity which denote contributions from the spin-orientation dependence ($\mathbf{n}$) and from the overlap of conduction basis states in different valleys: $\langle L_{1,\mathbf{k}_{\scriptscriptstyle{L}}} | L_{j,\mathbf{k}_{\scriptscriptstyle{L'}}} \rangle$. Eq.~(\ref{eq:overlap_inter}) implies that the ratio between spin and momentum relaxation rates due to intervalley scattering is about $\Delta_{L}^2/E_{g,u}^2$ (independent of the values of the wavevectors with respect to the valley centers). For intravalley scattering ($\mathbf{k}_{\scriptscriptstyle{L}}$=$\mathbf{k}_{\scriptscriptstyle{L}}$), on the other hand,  the basis functions are orthogonal: $\langle L_{\mu,\mathbf{k}_{\scriptscriptstyle{L}}}^n | L_{\gamma,\mathbf{k}_{\scriptscriptstyle{L'}}}^m \rangle = \delta_{\mu\gamma}\delta_{mn}$. As a result, the spin-flip overlap integral for electrons of the same valley reads
\begin{eqnarray}\label{eq:overlap_intra}
\!\!\!\!\!\! \mathcal{I}_a( \mathbf{k}',\Downarrow_{\scriptstyle{\Lambda}}\,;\,\mathbf{k},\Uparrow_{\scriptstyle{\Lambda}})  &=& \frac{2P^2}{E_{g,v}^2}\big( \gamma_2q_+K_{\scriptstyle{\Lambda}} + i\gamma_1q_-K_- \big),
\end{eqnarray}
where $\mathbf{q}=\mathbf{k}-\mathbf{k}'$, $2\mathbf{K}=\mathbf{k}+\mathbf{k}'$, and $X_{\pm}=X_w \pm iX_u$. The terms have quadratic wavevector dependence and they are proportional to the spin-orbit constants in Eq.~(\ref{eq:gamma_j}).\cite{footnote_gamma_j2} The overlap integral of other spin orientations ($\hat{\mathbf{n}}$$\,$$\neq$$\,$$\hat{\scriptstyle{\Lambda}}$) will be discussed in the next section.


\section{Intravelley spin relaxation}
\label{sec:V}

The power-law dependence of intravalley spin-flip matrix elements can be identified by their transformation properties under time reversal and space inversion operations. Yafet showed that spin-flip matrix elements due to scattering with long-wavelength acoustic phonons have a cubic (quadratic) wavevector dependence in Ge (Si).\cite{Yafet_1963}  In Appendix C, these important findings are generalized and it is shown that in Ge $\langle \mathbf{k}', \Downarrow \! |H_{\text{intra}}^{\lambda}| \mathbf{k}, \Uparrow \rangle$
scales with $K_{\ell}q_mq_n$ for scattering with acoustic phonon modes ($\lambda$$\,$=$\,$LA or TA) and with $K_{\ell}q_m$ for optical phonon modes ($\lambda$$\,$=$\,$LO or TO). $\mathbf{K}$ and $\mathbf{q}$ are, respectively, the average and difference of $\mathbf{k}$ and $\mathbf{k}'$. For intravalley scattering in Si the $\mathbf{K}$ dependence drops. We first explain this interesting difference.

From inspection of the wavevector dependence of intravalley spin flips in Ge ($Kq^i$), one sees that they are forbidden between opposite points with respect to the valley center ($\mathbf{K}=0$). This restriction on spin-flip transitions is a manifestation of time reversal symmetry. In silicon, $\mathbf{K}$-dependent scattering belongs to the intervalley $g$-process which involves transitions between two valleys on opposite sides of the same crystal axis.\cite{Li_PRL11,Song_arXiv12} Since in Ge the valley center is at the zone edge ($L$~point), this type of scattering occurs within a single valley. Its dependence on the wavevector components ($Kq^i$) amounts to the combined effects of intervalley $g$-process and intravalley scattering in Si ($K$ and $q^i$).

Beyond the power-law dependence, an analytical approach to derive accurate intravalley matrix elements requires a combination of $\mathbf{k}\cdot\mathbf{p}$, rigid-ion and group theories.\cite{Song_arXiv12} Because of the wavevector dependence of these matrix elements, one cannot invoke group theory alone to find their exact forms (as we did for zeroth-order intervalley spin flips). We employ a simpler approach than in Ref.~[\onlinecite{Song_arXiv12}] and describe the interaction with long-wavelength acoustic phonons by $H_{\text{intra}}^{\text{TA/LA}} = \Xi q$ where $\Xi$ is an effective deformation potential constant.\cite{Cardona_Book,Li_PRL11} This scalar form averages out the scattering angle dependence of the second-rank deformation potential tensor.\cite{Bir_Pikus_Book} We do not model the electron scattering with long-wavelength optical phonons since it is a weak effect in nonpolar semiconductors.\cite{Cardona_Book,footnote_optical}

We use selection rules of the $L$~point space group to construct the spin-flip matrix element from the overlap integral. The transformation property of the deformation potential tensor,  $L_3\!\!\!'\,\,\otimes L_3\!\!\!'\,\, = L_1+L_2+L_3$, implies that direct coupling of conduction and valence states is excluded because of their opposite parities ($L_1\otimes L_3\!\!\!'\,\, = L_3\!\!\!'\,\,$).  This tensor can, however, couple any of the basis states to themselves ($L_i\otimes L_i$). This behavior justifies the use of the spin-flip overlap integral. The resulting intravalley spin-flip matrix element in the $L_{111}$ valley is approximated by
\begin{eqnarray}
M_{intra}^{sf}(\mathbf{k}',\mathbf{k}; \mathbf{n})  \approx  \Xi q \mathcal{I}_a( \mathbf{k}',\Downarrow_{\mathbf{n}}\,;\,\mathbf{k},\Uparrow_{\mathbf{n}})\,. \label{eq:Mintra}
\end{eqnarray}
Following a straightforward procedure we find
\begin{eqnarray}
\mathcal{I}_a( \mathbf{k}',\Downarrow_{\mathbf{n}}\,;\,\mathbf{k}, \Uparrow_{\mathbf{n}}  ) = i\sin\vartheta A_t + A_l \cos^2\!\frac{\vartheta}{2} + A_l^{\ast} \sin^2\!\frac{\vartheta}{2} \,,\,\,\, \label{eq:overlap_intra_gen}
\end{eqnarray}
where $\cos\vartheta = \mathbf{n}\cdot\hat{\mathbf{\scriptstyle{\Lambda}}}$ and
\begin{eqnarray}
A_t &=& \frac{i}{2}\Big( \mathcal{I}_a( \mathbf{k}',\Uparrow_{\scriptstyle{\Lambda}}\,;\,\mathbf{k},\Uparrow_{\scriptstyle{\Lambda}}) - \mathcal{I}_a( \mathbf{k}',\Downarrow_{\scriptstyle{\Lambda}}\,;\,\mathbf{k},\Downarrow_{\scriptstyle{\Lambda}}) \Big) \nonumber \\
&=& \frac{2P^2}{E_{g,v}^2} \gamma_3  \left( \mathbf{K}\times\mathbf{q} \right)_{\scriptstyle{\Lambda}}, \nonumber \\
A_l &=& e^{-i\varphi} \mathcal{I}_a( \mathbf{k}',\Downarrow_{\scriptstyle{\Lambda}}\,;\,\mathbf{k},\Uparrow_{\scriptstyle{\Lambda}}) \nonumber \\ &=& \frac{2P^2}{E_{g,v}^2}\big( \gamma_2q_+K_{\scriptstyle{\Lambda}} + i\gamma_1q_-K_- \big)e^{-i\varphi}.
 \label{eq:Alt}
\end{eqnarray}
$\varphi$ is the azimuthal angle of $\mathbf{n}$ measured with respect to the $w$-axis in the $wu$-plane. Most importantly, $\gamma_3$ which incorporates the effect of the internal spin-orbit coupling in the valence band [Eq.~(\ref{eq:gammapm})] does not affect the spin-flip amplitude [Eq.~(\ref{eq:Mintra})] when the spin orientation is along the valley axis ($\vartheta = 0$). This effect will lead to a pronounced anisotropy in the intravalley spin lifetime.

It is not surprising that the overlap integral approach yields correct wavevector power-law dependence [substituting Eqs.~(\ref{eq:overlap_intra_gen})-(\ref{eq:Alt}) into Eq.~(\ref{eq:Mintra})]. The space inversion and time reversal symmetries are respected by the Hamiltonian whose eigenvectors were used to calculate the intravalley spin-flip overlap integral. These symmetries also lead to the so-called Elliott-Yafet cancelation of all terms up to quadratic order in $\mathbf{q}$.\cite{Song_arXiv12,Yafet_1963} In fact, since the Hamiltonian respects all other symmetries of the $L$~point space group, the intravalley matrix element shows other selection rules.\cite{footnote_scattering_symmetries} From Eq.~(\ref{eq:Alt}) we see, for example, that a spin-flip is forbidden when the electron is scattered along the valley axis (i.e., $q_w$=$q_u$=0, $q_{\scriptscriptstyle{\Lambda}}$$\neq$0). This constraint is understood by the symmetry of the vector-type coupling with the valence states ($L_1\otimes L_3\!\!\!'\,\,=L_3\!\!\!'\,\,$). As mentioned, this coupling involves the two transverse components ($\hat{\mathbf{w}}$ and $\hat{\mathbf{u}}$)  with respect to the valley axis ($\hat{\mathbf{\scriptstyle{\Lambda}}}$).

We calculate the spin lifetime in the $L_{111}$ valley due to electron scattering with long-wavelength acoustic phonon modes. This intravalley process dominates the spin relaxation under conditions of \textit{$[111]$~strain}.\cite{Tang_PRB12} For sufficient uniaxial compressive strain along this direction ($\sim1\%$), one of the valleys is significantly lowered in energy and the intervalley scattering is quenched. Then, phonon-induced intravalley spin-flips can dictate the spin relaxation of conduction electrons if scattering from impurities is negligible (non-degenerate doping). To get an analytical expression of the intravalley spin lifetime, the phonon energy is approximated by $\Omega_{\scriptscriptstyle{AC}}(\mathbf{q})=~\!\!\hbar v_{\scriptscriptstyle{AC}}q$ where $v_{\scriptscriptstyle{AC}} \simeq 3.5\cdot10^5$~cm/sec is the speed of acoustic phonons in Ge. We also make use of the long-wavelength limit and approximate the acoustic phonon population by $k_BT/\Omega_{\scriptscriptstyle{AC}}(\mathbf{q}) \gg 1$. Then by considering a Boltzman distribution of  electrons and substituting Eqs.~(\ref{eq:Mintra})-(\ref{eq:Alt}) into Eq.~(\ref{eq:ts_phonon}) one gets
\begin{eqnarray}
\frac{1}{\tau_{s,intra}} = \frac{\gamma_{\scriptscriptstyle{3}}^2}{\tau_{\scriptscriptstyle{0}}} \left(\frac{k_BT}{U_0}\right)^{\!\frac{7}{2}} \! \left[ \sin^2\vartheta + (1+\cos^2\vartheta)\beta  \right], \label{eq:ts_intra}
\end{eqnarray}
where $U_{0}$$\,$=$\,$25.8~meV is the room-temperature thermal energy. $\beta$$\,$$\approx$$\,$0.12 and $\tau_{\scriptscriptstyle{0}}$$\,$$\approx$$\,$0.3~ns are expressed by
\begin{eqnarray}
\beta &=& \frac{2m_l\gamma_2^2+3m_t\gamma_1^2}{5m_t\gamma_3^2}, \label{eq:anisotropy_factor} \\
\frac{1}{\tau_{\scriptscriptstyle{0}}} &=& \frac{1024}{3} \left( \!1- \frac{m_t}{m_t^{\ast}} \right)^{\!\!2} \! \frac{\Xi^2}{E_{g,v}^2} \!\left( \frac{m_d}{2\pi} \right)^{\frac{3}{2}} \frac{U_{0}^{\frac{7}{2}}}{\hbar^4\varrho v_{\scriptscriptstyle{AC}}^2}.\,\,\,
\label{eq:t0_intra}
\end{eqnarray}
In accord with momentum scattering, we have used a value of $\Xi$$\,$=$\,$7.5~eV for the deformation potential constant.\cite{Cardona_Book} The anisotropy in the intravalley spin relaxation is evident [square bracket term in Eq.(\ref{eq:ts_intra})]. Our analysis shows that the lifetime is the longest for spin orientation along the valley (and strain) axis where $\vartheta$$\,$=$\,$0. It drops by nearly a factor of 5 when the spin is oriented in the perpendicular plane ($\vartheta$$\,$=$\,$$\pi$/2). At room temperature, this change amounts to reducing the intravalley spin lifetime from $\sim$700~ns to $\sim$150~ns. These extremely long timescales are a consequence of the space inversion symmetry and the position of the valley center in the edge of the Brillouin zone.

The temperature dependence of the intravalley spin lifetime is shown by the dash line in Fig.~\ref{fig:ts}(b) for spin orientation along the $z$-axis [assigning $\cos^2\vartheta$$\,$=$\,$1/3 in Eq.~(\ref{eq:ts_intra})]. In this spin orientation, the intravalley spin relaxation rate is equivalent in all four valleys. Figure~\ref{fig:ts}(b) also shows that in unstrained bulk Ge, the spin lifetime of conduction electrons due to intravalley scattering with acoustic phonons is two orders of magnitude longer than the intervalley spin lifetime at room temperature. In addition, at very low temperatures the intrinsic spin lifetime reaches timescales of one second. Therefore, the phonon-induced spin relaxation is likely to be readily masked at low temperatures by localization effects on residual impurities (e.g., hyperfine interactions and Raman processes).\cite{Song_arXiv12} Finally, the square markers in  Fig.~\ref{fig:ts}(b) show results of rigorous numerical calculations following the procedure in Ref.~[\onlinecite{Cheng_PRL10}]. Evidently, the overlap integral analytical approach provides rather accurate results and yet it clearly explains the underlying physics.

Before concluding this part, we compare three aspects of the intravalley spin relaxation in Si and Ge. First, the overlap integral approach is valid in Ge due to the relatively large separation of the non-degenerate conduction band from other valence and conduction bands. In Si, on the other hand, the intravalley spin relaxation is affected by the proximity of the conduction bands where the off-diagonal terms of the deformation potential play a key role.\cite{Li_PRL11,Song_arXiv12} Second, along the $\Delta$-symmetry axis which is relevant in Si, the spin-orbit coupling does not lift the energy degeneracy between the upper pair of valence bands. As a result, the intravalley spin relaxation is not affected by the internal spin-orbit coupling in the valence band and the anisotropy is weaker in Si reaching a factor of two.\cite{Song_arXiv12} Finally, the intravalley spin relaxation rate exceeds the intervalley rate below 50~K in Si,\cite{Huang_PRL07} and below 20~K in Ge. Reasons for the difference are the larger energy of zone-edge phonons in Si and the $T^{5/2}$ rather than $T^{7/2}$ temperature dependence of its intravalley process.

\section{summary}
\label{sec:VI}
We have presented various origins that limit the intrinsic spin lifetime of conduction electrons in Ge. In unstrained bulk Ge and at T$\,$$>$$\,$20~K, the intrinsic spin lifetime is limited by intervalley electron scattering with zone-edge phonon modes of $X_1$ and $X_4$ symmetries (reaching $\sim$1~ns at 300~K). This spin lifetime is governed by the coupling with the upper conduction bands and its temperature dependence is set by the thermal population of the zone-edge phonons (with energies of about 30~meV).  By analyzing the crystal and time reversal symmetries in the multivalley conduction band, we have found the spin orientation dependence of the dominant intervalley spin-flip processes. This dependence allowed us to quantify the change in the intervalley spin lifetime when varying the spin orientation under various stress configurations [Eqs.~(\ref{eq:ts_inter})-(\ref{eq:A_110})].

We have derived a spin-dependent $\mathbf{k}$$\cdot$$\mathbf{p}$ Hamiltonian model in the vicinity of the zone-edge $L$~point [Eqs.~(\ref{eq:10by10})-(\ref{eq:H_uv})]. This compact model provides a lucid picture of the spin-orbit coupling effects in Ge. Similar to using the Kane model in zinc-blend semiconductors,\cite{Kane_JPCS57} the compact $L$~point Hamiltonian has implications beyond derivation of spin-flip matrix elements. For example, by employing a plane-wave expansion along confined directions in nanostructures, this Hamiltonian model can be used to study spin-dependent properties in Ge nanostructures. Together with Si related theories,\cite{Li_PRL11,Song_arXiv12} one can also investigate spin properties in SiGe alloys.

Using the eigenvectors of the Hamiltonian matrix, we have derived forms of the spin-flip matrix elements due to intravalley scattering with long-wavelength acoustic phonons [Eqs.~(\ref{eq:Mintra})-(\ref{eq:Alt})]. The intravalley spin lifetime is found to be two order of magnitude longer than the intervalley spin lifetime.  As such, intravalley spin-flips affect the overall spin relaxation only when quenching the intervalley spin relaxation (e.g., by application of a uniaxial compressive stress along the [$111$] crystallographic axis).\cite{Tang_PRB12} Beyond the $T^{7/2}$ temperature dependence of the intravalley spin relaxation, we have also quantified its dependence on the spin orientation [Eq.~(\ref{eq:ts_intra})]. The anisotropy of the intravalley spin relaxation results in a remarkably long spin lifetime (nearly 1~$\mu$s at room-temperature) when the spin is oriented along the valley (and strain) axis.  The relatively large anisotropy of the intravalley spin relaxation was explained by the coupling with the internal spin-orbit interaction in the valence band.

We have elucidated the differences in the spin relaxation of bulk Si and Ge crystals. While both materials have a diamond-crystal structure, in Ge the valley center is located at the edge of the Brillouin zone ($L$ point) and the lowest conduction band is well separated from other bands. These properties lead to a very long intravalley spin lifetime in Ge with a cubic power-law dependence of intravalley spin-flips on wavevector components. This cubic dependence is also expected to be larger than in graphene where unlike Ge but similar to Si, the time-reversal operation couples states in inequivalent valleys. Therefore, in spite of being heavier than Si and carbon, non-degenerate and strained bulk Ge is a very promising material choice for implementing spintronic devices.\cite{Dery_PRB06b,Cywinsky_APL06,Fabian_APS_07,De_Sousa_PRB09}

This work is supported by AFOSR Contract No. FA9550-09-1-0493 and by NSF Contract No. ECCS-0824075.


\appendix
\section{Derivation of the selection rules for intervalley spin-flip transition}
We first focus on scattering between the $L_{111}$ and $L_{11\overline{1}}$ valley centers [$\mathbf{k}_L=(1,1,1)/2$ and $\mathbf{k}_{Lt}=(1,1,-1)/2$]. Generalization to other valley centers is made at the end of the Appendix.

The selection rules connecting $L$ and $L_t$ points involve common symmetry operations of the little groups at $\mathbf{k}_L$, $-\mathbf{k}_{Lt}$ and $\mathbf{q}_X=\mathbf{k}_{Lt}-\mathbf{k}_L$,
\begin{eqnarray}
g_c\in&\{& (\epsilon|0), (\bar{\epsilon}|0),(\delta_{2x\bar{y}}|\tau), (\bar{\delta}_{2x\bar{y}}|\tau),\nonumber\\
&&(i|\tau), (\bar{i}|\tau), (\rho_{x\bar{y}}|0),  (\bar{\rho}_{x\bar{y}}|0) \,\,\,\,\}.
\end{eqnarray}
They also involve operations that switch between $\mathbf{k}_L$ and $-\mathbf{k}_{Lt}$,
\begin{eqnarray}
g_e \in &\{&(\delta_{2z}|0), (\bar{\delta}_{2z}|0), (\rho_{z}|\tau),  (\bar{\rho}_{z}|\tau), \nonumber\\
&&(\rho_{xy}|0),  (\bar{\rho}_{xy}|0)\}, (\delta_{2xy}|\tau), (\bar{\delta}_{2xy}|\tau)\,\,\,\,\}.
\end{eqnarray}
The bar over operations denotes an additional $2\pi$ rotation (in double group notation). Table~\ref{tab:character} lists the characters of the nontrivial operations. By considering these operations and time reversal symmetry, the number of independent nonvanishing matrix elements for each of the zone-edge phonon symmetries in diamond-crystal structures ($X_1$, $X_3$, $X_4$) is given by
\begin{eqnarray}
\mathcal{N}_{X_i}=\frac{1}{2h_0}\left[\sum\limits_{g_c} \chi^{\mathbf{-k}_{Lt}}_{L^+_{6t}}(g_c) \chi^{\mathbf{k}_L}_{L^+_6}(g_c) \chi^{\mathbf{q}_X}_{X_i}(g_c)\right. \nonumber\\
\left.- \sum\limits_{g_e} \chi^{\mathbf{k}_L}_{L^+_6}(g^2_e) \chi^{\mathbf{q}_X}_{X_i}(g_e) \right], \label{eq:indep_num}
\end{eqnarray}
where $h_0=8$ is the number of $g_c$ or $g_e$ operations and $\chi_{L^+_{6(t)}}=\chi_{L_{1(t)}} \times \chi_{1/2}$. The second sum in Eq.~(\ref{eq:indep_num}) denotes the effect of time reversal symmetry and the minus sign takes into account the parity from the spinor basis and interaction $H_{\rm{ep}}$ (see, Ref.~[\onlinecite{Bir_Pikus_Book}] for more details).  By straightforwardly plugging the characters of Table~\ref{tab:character} into Eq.~(\ref{eq:indep_num}) one finds the general selection rule of Eq. (\ref{eq:phonons}).

\begin{table}
\caption{\label{tab:character}
Non-trivial relevant IR characters and matrices in a intervalley scattering between $\mathbf{k}_L$ and $\mathbf{k}_{Lt}$ valleys. For ID IR $L_{1(t)}$ and 2D IR $X_3$, only characters are used and shown. $I$ and $\sigma_x$ used in 2D IR $X_1$ and $X_4$ are the $2\times2$ identity matrix and Pauli matrix.   These matrices  are based our choice of basis states. The final results do not depend on this specific choice since the two phonon modes belonging to each IR are degenerate. $\chi^{-\mathbf{k}_{Lt}}_{L_t}=\chi^{\mathbf{k}_{Lt}}_{L_t}$. Also shown is the effect of exchange operations on $L$ star. Basis states in $D_{1/2}$ is along $\pm z$ in spin space.  }
\renewcommand{\arraystretch}{2.5}
\tabcolsep=0.2 cm
\begin{tabular}{c|c|c|c|c|c|c}
\hline \hline
                             & $X_1$      & $X_3$     &$X_4$      &  $L_1$      & $L_{1t}$      & $D_{1/2}$\\
               \hline
 $(\delta_{2x\bar{y}}|\tau)$ & $\sigma_x$ & $-I$      &  I        &  1          & -1            &  $\displaystyle
 e^{\mbox{-}\tfrac{3\pi i}{4}}\!\! \left(\!\!\!\renewcommand{\arraystretch}{1.0}\begin{array}{cc}0&i\\1&0\end{array}\!\!\!\right)$\\
 \hline
 $(i|\tau)$                  & $\sigma_x$            & 0       &  $\sigma_x$    &  1          & -1            & $\displaystyle
  \left(\!\!\!\renewcommand{\arraystretch}{1.0}\begin{array}{cc}1 & 0\\ 0 & 1\end{array}\!\!\!\right)$\\
\hline
 $(\rho_{x\bar{y}}|0)$        & $I$           & 0       &  $\sigma_x$  &  1          & 1            & $\displaystyle
 e^{\mbox{-}\tfrac{3\pi i}{4}}\!\!  \left(\!\!\!\renewcommand{\arraystretch}{1.0}\begin{array}{cc}0 & i\\ 1 & 0\end{array}\!\!\!\right)$\\
 \hline
 $(\delta_{2z}|0)$            & $I$           & -2       &  -$I$        & \multicolumn{2}{|c|}{$\mathbf{k}_L\!\!\!\leftrightarrow\!\! \mbox{-}\mathbf{k}_{Lt}$\!}           & $\displaystyle
 \!\!  \left(\!\!\!\renewcommand{\arraystretch}{1.0}\begin{array}{cc}-i & 0\\ 0 & i\end{array}\!\!\!\right)$\\
 \hline
 $(\rho_{z}|\tau)$            & $\sigma_x$           & 0        &  -$\sigma_x$      & \multicolumn{2}{|c|}{$\mathbf{k}_L\!\!\!\leftrightarrow\!\! \mbox{-}\mathbf{k}_{Lt}$\!}           & $\displaystyle
 \!\!  \left(\!\!\!\renewcommand{\arraystretch}{1.0}\begin{array}{cc}-i & 0\\ 0 & i\end{array}\!\!\!\right)$\\
 \hline
  $(\rho_{xy}|0)$            & $I$           & 0        & -$\sigma_x$       & \multicolumn{2}{|c|}{$\mathbf{k}_L\!\!\!\leftrightarrow\!\! \mbox{-}\mathbf{k}_{Lt}$\!}           & $\displaystyle
 e^{\mbox{-}\tfrac{3\pi i}{4}}\!\!  \left(\!\!\!\renewcommand{\arraystretch}{1.0}\begin{array}{cc}0 & 1\\ i & 0\end{array}\!\!\!\right)$\\
  \hline
  $(\delta_{2xy}|\tau)$      & $\sigma_x$   & -2   &  $I$      & \multicolumn{2}{|c|}{$\mathbf{k}_L\!\!\!\leftrightarrow\!\! \mbox{-}\mathbf{k}_{Lt}$\!}           & $\displaystyle
 e^{\mbox{-}\tfrac{3\pi i}{4}}\!\!  \left(\!\!\!\renewcommand{\arraystretch}{1.0}\begin{array}{cc}0 & 1\\ i & 0\end{array}\!\!\!\right)$\\
\hline\hline
\end{tabular}
\end{table}

Our aim is to express interaction matrix elements $\langle \mathbf{k}_{Lt},\mathbf{s}_2 |H_{X_i}| \mathbf{k}_L,\mathbf{s}_1 \rangle$ between specific spin species in terms of $\mathcal{N}_{X_i}$ independent constants. In order to identify these scattering constants, we connect different matrix elements via appropriate symmetry operations.

First, by time reversal and space inversion symmetries of diamond-crystal structures we can write
\begin{eqnarray}
\langle \mathbf{k}_{Lt}, \Uparrow |H_{X_i}| \mathbf{k}_L, \Uparrow \rangle & = & \langle \mathbf{k}_{Lt}, \Downarrow |{H}_{X_i}| \mathbf{k}_L, \Downarrow \rangle^*, \label{eq:TS_m}\\
\langle \mathbf{k}_{Lt}, \Downarrow |{H}_{X_i}| \mathbf{k}_L, \Uparrow \rangle & = & -\langle \mathbf{k}_{Lt}, \Uparrow |{H}_{X_i}| \mathbf{k}_L, \Downarrow \rangle^*.\label{eq:TS_s}
\end{eqnarray}
These identities hold for all phonons and possible spin orientations. The minus sign in Eq. (\ref{eq:TS_s}) roots from the Pauli matrix $\sigma_y$ in the time reversal operator $\hat{T}=\hat{K}\sigma_y$, where $\hat{K}$ is the complex conjugate operator.

We first study the case $\mathbf{n}\|\mathbf{z}$, where $\mathbf{n}$ is the spin orientation. For $X_1$, the $(\rho_{x\bar{y}}|0)$ operation equates spin-conserving  transition to itself (seen from the IR matrices of $X_1$ and $D_{L^+_{6(t)}}=D_{L_{1(t)}} \times D_{1/2}$ in Table~\ref{tab:character}),
\begin{eqnarray}
\langle \mathbf{k}_{Lt}, \Uparrow |H_{X^{a(b)}_1}| \mathbf{k}_L, \Uparrow \rangle  \stackrel{(\rho_{x\bar{y}}|0)}{=}  \langle \mathbf{k}_{Lt}, \Downarrow |H_{X^{a(b)}_1}| \mathbf{k}_L, \Downarrow \rangle, \label{eq:X1_m_1}
\end{eqnarray}
where two $X_1$ basis states are denoted as $X^a_1$ and $X^b_1$. This choice is arbitrary and will not affect the final results due to the 2-fold degeneracy of the phonon modes.  Eqs.~(\ref{eq:TS_m}) and (\ref{eq:X1_m_1}) require the matrix elements of each of the $X_1$ phonon branches to be a real number. From Table~\ref{tab:character}, one can also find $(i|\tau)$ relates the matrix elements of the two degenerate modes by a minus sign,
\begin{eqnarray}
\langle \mathbf{k}_{Lt}, \Uparrow |H_{X^{a(b)}_1}| \mathbf{k}_L, \Uparrow \rangle \stackrel{(i|\tau)}{=} -\langle \mathbf{k}_{Lt}, \Uparrow |H_{X^{b(a)}_1}| \mathbf{k}_L, \Uparrow \rangle. \label{eq:X1_m_2}
\end{eqnarray}
With this additional information, a real number $D_{X_1,m}$ could be assigned such that
\begin{eqnarray}
\langle \mathbf{k}_{Lt}, \Uparrow\!\! |H_{X^{a}_1}\!| \mathbf{k}_L, \Uparrow \rangle \! =\!  -\langle \mathbf{k}_{Lt}, \Uparrow \!\!|H_{X^{b}_1}| \mathbf{k}_L, \Uparrow \rangle\!=\!D_{X_1,m}, \label{eq:X1_m}
\end{eqnarray}
Other operations do not give further information on these matrix elements.

\begin{table*}
\renewcommand{\arraystretch}{1.5}
\tabcolsep=0.7cm
\caption{\label{tab:M_Sn}
$\sum_{j=a,b}|M_{X^{j}_i} (\mathbf{k}_L,\Uparrow_\mathbf{n}; \mathbf{k}_{Lt},\Downarrow_\mathbf{n})/ D_{X_i}|^2$ for intervalley spin flips between $L_{111}$ and $L_{11\bar{1}}$ valleys. For each of the non-vanishing modes, $X_i$, the relative amplitude is provided for spin orientation ($\mathbf{n})$ along any of the inequivalent high-symmetry crystal directions. Results between other valleys can all be obtained by trivial rotation transformation.
}
\begin{tabular}{cccccccc}
\hline \hline
$\mathbf{n}$    &   $[0\;0\;1]$  & $[1\;0\;0]$ & $[1\;1\;0]$ & $[1\;\bar{1}\;0]$ &  $[1\;0\;1]$ &  $[1\;1\;1]$  & $[1\;\bar{1}\;1]$     \\ \hline
$X_1$           &       4        &      2      &        4    &      0            &      3       &       4       &       4/3             \\
$X_4$           &       0        &      2      &        2    &      2            &      1       &       4/3     &       4/3             \\ \hline \hline
\end{tabular}
\end{table*}

With the same operations, the result for spin-flip transition is
\begin{eqnarray}
\!\!\!\!\!\!\!\!\langle \mathbf{k}_{Lt}, \Downarrow\!\! |H_{X^{a(b)}_1}\!| \mathbf{k}_L, \Uparrow \rangle  \!\!\!\!&\stackrel{(\rho_{x\bar{y}}|0)}{=}&\!\!\!\!  -i\langle \mathbf{k}_{Lt}, \Uparrow \!\!|H_{X^{a(b)}_1}\!| \mathbf{k}_L, \Downarrow \rangle, \label{eq:X1_s_1}\\
\!\!\!\!\!\!\!\!\langle \mathbf{k}_{Lt}, \Downarrow \!\!|H_{X^{a(b)}_1}\!| \mathbf{k}_L, \Uparrow \rangle  \!\!\!\!&\stackrel{(i|\tau)}{=}&\!\!\!\!  -\langle \mathbf{k}_{Lt}, \Downarrow \!\!|H_{X^{a(b)}_1}\!| \mathbf{k}_L, \Uparrow \rangle. \label{eq:X1_s_2}
\end{eqnarray}
Together with Eq.~(\ref{eq:TS_s}), we can assign a real number $D_{X_1,s}$ such that
\begin{eqnarray}
\langle \mathbf{k}_{Lt}, \Downarrow |H_{X^{a}_1}| \mathbf{k}_L, \Uparrow \rangle  &=&  -\langle \mathbf{k}_{Lt}, \Downarrow |H_{X^{b}_1}| \mathbf{k}_L, \Uparrow \rangle\nonumber\\
&=&(1+i)D_{X_1,s}. \label{eq:X1_s}
\end{eqnarray}

Next we analyze matrix elements due to $X_4$ modes, where there is only one independent scattering constant. From Table~\ref{tab:character}, the operations $(\delta_{2x\bar{y}}|\tau)$ and $(i|\tau)$ give relations for spin-conserving transitions
\begin{eqnarray}
\!\!\!\!\!\!\!\!\!\!\!\!\langle \mathbf{k}_{Lt}, \Uparrow\!\! |H_{X^{a(b)}_4}| \mathbf{k}_L, \Uparrow \rangle \!\!\!\!&\stackrel{(\delta_{2x\bar{y}}|\tau)}{=}&\!\!\!\! -\langle \mathbf{k}_{Lt}, \Downarrow \!\!|H_{X^{a(b)}_4}| \mathbf{k}_L, \Downarrow \rangle, \label{eq:X4_m_1}\\
\!\!\!\!\!\!\!\!\!\!\!\!\langle \mathbf{k}_{Lt}, \Uparrow \!\!|H_{X^{a(b)}_4}\!| \mathbf{k}_L, \Uparrow \rangle \!\!\!\!&\stackrel{(i|\tau)}{=}&\!\!\!\! -\langle \mathbf{k}_{Lt}, \Uparrow \!\!|H_{X^{b(a)}_4}\!| \mathbf{k}_L, \Uparrow \rangle. \label{eq:X4_m_2}
\end{eqnarray}
Together with Eq.~(\ref{eq:TS_m}), a real number $D_{X_4,s}$ could be assigned
\begin{eqnarray}
\!\!\!\!\!\!\!\langle \mathbf{k}_{Lt}, \Uparrow\!\! |H_{X^{a}_4}\!| \mathbf{k}_L, \Uparrow \rangle \! =\!  -\langle \mathbf{k}_{Lt}, \Uparrow\!\! |H_{X^{b}_4}\!| \mathbf{k}_L, \Uparrow \rangle\!=\!i D_{X_4,s}. \label{eq:X4_m}
\end{eqnarray}

For spin-flip transitions, the exchange operation $(\delta_{2z}|0)$ together with the general time reversal operation connect the matrix elements to their negatives,
\begin{eqnarray}
\langle \mathbf{k}_{Lt}, \Downarrow |H_{X^{a(b)}_4}| \mathbf{k}_L, \Uparrow \rangle
\stackrel{(\delta_{2z}|0)}{=} &&\\\langle \mbox{-}\mathbf{k}_{L}, \Downarrow |H_{X^{a(b)}_4}| \mbox{-}\mathbf{k}_{Lt}, \Uparrow \rangle
&\stackrel{\rm{TR}}{=} &-\langle \mathbf{k}_{Lt}, \Downarrow |H_{X^{a(b)}_4}| \mathbf{k}_{L}, \Uparrow \rangle,\nonumber \label{eq:X4_s}
\end{eqnarray}
where the time reversal operation sends electron states to their Kramers conjugate, and keeps the electron-phonon interaction. Thus spin-flip matrix elements duo to $X_4$ phonon modes vanish.

Therefore,  with spin direction along $z$, the scattering matrices from $\mathbf{k}_L$ to $\mathbf{k}_{Lt}$ for relevant phonon modes are
\begin{eqnarray}
\!\!\!\!\!\!\!\!\!H_{X^{a}_1}\!=\!-H_{X^{b}_1}\!\!\!&=&\!\!\!\renewcommand{\arraystretch}{1.6}\left(\!\!\begin{array}{cc} \!D_{X_1,m}\!\!&\!(-1+i)D_{X_1,s}\\(1+i)D_{X_1,s}\!\!\! & \!\! D_{X_1,m}\!\end{array}\!\!\right)\!,\label{eq:H_X1}\\
\!\!\!\!\!\!\!\!\!H_{X^{a}_4}\!=\!-H_{X^{b}_4}\!\!&=&\!\!\renewcommand{\arraystretch}{1.6}\left(\!\!\begin{array}{cc} \!i D_{X_4,s}\!\!&\!0\\0\!\!\! & \!\! -i D_{X_4,s}\!\! \end{array}\!\!\right),\label{eq:H_X4}
\end{eqnarray}
where Eqs.~(\ref{eq:TS_m}) and (\ref{eq:TS_s}) are used to get two other elements in each matrix. Eqs.~(\ref{eq:H_X1}) and (\ref{eq:H_X4}) indicate that in this specific case, $X_1$ is allowed for both spin-conserving and spin-flip transitions, while $X_4$ is only allowed for spin-conserving transition, originate from the $z$ component of the spin-orbit interaction [$\propto (\nabla V\times p)_z\sigma_z$] that does not flip spin.

Next we extend the analysis to arbitrary spin orientation, which leads to the anisotropy of spin relaxation processes and enables a direct comparison to a wide range of spin injection experiments.

The spin orientation ($\mathbf{n})$ is defined in terms of polar ($\theta$) and azimuthal angles ($\phi$) with respect to the $+z$ and $+x$ directions. The new spin states relate to the original ones by an `active' rotation matrix in spin sub-space,
\begin{eqnarray}
\displaystyle\exp\left(\frac{i\bm\sigma \!\cdot\! \hat{\bm{\omega}} \,\theta}{2}\right)= \left(\renewcommand{\arraystretch}{2.0} \begin{array}{cc} \displaystyle\cos{\frac{\theta}{2}} & \displaystyle-\sin\!{\frac{\theta}{2}}\,e^{-i\phi}\\  \displaystyle\sin\!{\frac{\theta}{2}}\,e^{i\phi} & \displaystyle\cos{\frac{\theta}{2}}\end{array}\right),\label{eq:rot_matrix}
\end{eqnarray}
where $\hat{\bm{\omega}}=\hat{\mathbf{n}} \times \hat{\mathbf{z}}/|\hat{\mathbf{n}} \times \hat{\mathbf{z}}|$ is the unit vector along the rotation axis. The new spin states follow
\begin{eqnarray}
\!\!\!\!\!\!\!\!|\mathbf{k}_{\scriptscriptstyle{L}}, \Uparrow_\mathbf{n} \rangle &=&\cos\frac{\theta}{2}|\mathbf{k}_{\scriptscriptstyle{L}}, \Uparrow_z \rangle +\sin\frac{\theta}{2} e^{i\phi}|\mathbf{k}{\scriptscriptstyle{L}}, \Downarrow_z \rangle, \,\,\,\,\,\,\\
\!\!\!\!\!\!\!\! |\mathbf{k}_{\scriptscriptstyle{L}}, \Downarrow_\mathbf{n} \rangle &=&-\sin\frac{\theta}{2} e^{-i\phi}|\mathbf{k}_{\scriptscriptstyle{L}}, \Uparrow_z \rangle + \cos\frac{\theta}{2}|\mathbf{k}{\scriptscriptstyle{L}}, \Downarrow_z \rangle, \,\,\,\,\,\,
\end{eqnarray}
while the new scattering matrices from $\mathbf{k}_L$ to $\mathbf{k}_{Lt}$ are readily obtained by applying the rotation operator of Eq.~(\ref{eq:rot_matrix}) on the matrices Eqs. (\ref{eq:H_X1}) and (\ref{eq:H_X4}). The new spin-flip matrix elements are
\begin{eqnarray}
\!\!\!\!\!\!\!\!\!\!\!\!\!\!\!\!\!\!&&\langle \mathbf{k}_{Lt}, \Downarrow_\mathbf{n}\! |{H}_{X^a_1}| \mathbf{k}_L, \Uparrow_\mathbf{n} \rangle =
-\langle \mathbf{k}_{Lt}, \Downarrow_\mathbf{n}\! |{H}_{X^b_1}| \mathbf{k}_L, \Uparrow_\mathbf{n} \rangle \label{eq:Ms_X1_Sn} \\
\!\!\!\!\!\!\!\!\!\!\!\!\!\!\!\!\!\!&&= \left[(1+i)\cos^2\frac{\theta}{2} +(1-i)  \sin^2\frac{\theta}{2} e^{2i\phi}\right]  D_{X_1,s}\nonumber,
\\
\!\!\!\!\!\!\!\!\!\!\!\!\!\!\!\!\!\!&&\langle \mathbf{k}_{Lt}, \Downarrow_\mathbf{n}\! |{H}_{X^a_4}| \mathbf{k}_L, \Uparrow_\mathbf{n} \rangle =-\langle \mathbf{k}_{Lt}, \Downarrow_\mathbf{n}\! |{H}_{X^b_4}| \mathbf{k}_L, \Uparrow_\mathbf{n} \rangle  \!\! \label{eq:Ms_X4_Sn} \\
\!\!\!\!\!\!\!\!\!\!\!\!\!\!\!\!\!\!&&= i\sin\theta e^{i\phi} D_{X_4,s}.\nonumber
\end{eqnarray}
Summing the square amplitudes of the two branches leads to Eq. (\ref{eq:Ms}) in the paper. Table~\ref{tab:M_Sn} lists the relative magnitudes of the squared spin-flip matrix elements for $\mathbf{n}$ along several inequivalent high-symmetry directions of the crystal.


The matrix elements are determined by the relevant directions of the spin orientation and the valley-to-valley configurations. For configurations other than $L_{111}~\leftrightarrow~L_{11\overline{1}}$, the matrix elements could be obtained from Eq.~(\ref{eq:Ms}) by coordinate transformations. If we rewrite Eq.~(\ref{eq:Ms}) in the form of the projections of $\mathbf{n}$ on ${x,y,z}$ axes as
\begin{eqnarray}
\label{eq:Ms_xyz}
&&\sum_{i=1,2} |\langle \mathbf{k}_{Lt}, \Downarrow\! |{H}_{X^i_j}| \mathbf{k}_L, \Uparrow \rangle|^2 \\
=&&\left\{\begin{array}{ll}2D_{X_1,s}^2(1+\hat{z}^2 + 2\hat{x}\hat{y}),&\mbox{ if }
j = 1, \\ 2D_{X_4,s}^2 (1-\hat{z}^2),  & \mbox{ if } j=4,\end{array}\right. \nonumber
\end{eqnarray}
then in other valley-to-valley configurations, for example, $L_{\overline{1}11}\leftrightarrow L_{11\overline{1}}$, the matrix elements are just interchange $\{\hat{x},\hat{y},\hat{z}\}$ of Eq. (\ref{eq:Ms_xyz}) into $\{\hat{x},-\hat{z},\hat{y}\}$. Results of all possible configurations are listed in Table~\ref{tab:M_SL} of the paper.

\section{Derivation of the spin-dependent $L$-point Hamiltonian and the eigenstates}
We use the method of invariants to derive the Hamiltonian. The general procedure is: 1. Figuring out the two IRs of the coupling matrix; 2. Decomposing the direct product of these two IRs into a sum of IR(s); 3. According to this decomposition, associating invariant components and matrices to construct the Hamiltonian. These invariant components and matrices are obtained by applying the symmetry operators on the components of the perturbation and the chosen basis states, respectively. Associating the invariants to IRs is then carried by examining the resulting transformation. Table~\ref{tab:table_of_inv} lists these invariant components and matrices of the $L_{111}$~point. From this table, the constructions of $H_{ij}$ in Eqs.~(\ref{eq:H_cc})-(\ref{eq:H_uv}) are straightforward. Parameter constants of these Hamiltonian blocks are provided in Table~\ref{tab:params}.

\begin{table}[h]
\renewcommand{\arraystretch}{2}
\tabcolsep=0.1cm
\caption{\label{tab:table_of_inv}
Relevant invariant components and matrices of the $L_{111}$-point.
}
\begin{tabular}{c|c|c}
\hline \hline
IRs & Invariant components (111) & Invariant matrices\\ \hline
$L_1$ &  $k^2$, $-\frac{1}{2}(k_u^2+k_v^2)+k_w^2$ & $1$, $I$ \\\hline
$L_2$ &  $\sigma_w$ &  $\rho_y$  \\\hline
$L_3$ &  $\{\sigma_u,\sigma_v \}$ &   $ \{[0,1],[1,0]\},\{\rho_x,\rho_z\}$ \\\hline
$L_3'$ &  $\{-k_v,k_u \}$, $\{(\mathbf{k}\times\mbox{\boldmath$\sigma$})_u,(\mathbf{k}\times\mbox{\boldmath$\sigma$})_v\}$& $ \{[0,1],[1,0]\}$\\
\hline\hline
\end{tabular}
\end{table}

\begin{table}[h]
\caption{\label{tab:params}
Parameters of the $L$~point Hamiltonian [Eqs.~(\ref{eq:H_cc})-(\ref{eq:H_uv})] for bulk germanium following a spin-dependent empirical pseudopotential model. $m_0$ denotes the free electron mass.}
\renewcommand{\arraystretch}{1.2}
\tabcolsep=0.2 cm
\begin{tabular}{clc|clc|cl}
\hline \hline 
 $E_{g,u}$      & 2.2    & eV     &  $P$       & 9    & eV$\cdot$$\AA$   &  $m_t^{\ast}$        & 0.17$m_0$ \\
 $E_{g,v}$      & 3.3    & eV     &  $P_1$     & 7    & eV$\cdot$$\AA$   &  $m_l^{\ast}$        & 1.60$m_0$  \\
 $\Delta_L$     & 0.027  & eV     &  $P_2$     & 1.8  & eV$\cdot$$\AA$   &  $m_{t,3}^{\ast}$    & 1.2$m_0$    \\
 $\Delta_3$     & 0.022  & eV     &  $\alpha$  & 0.04 & eV$\cdot$$\AA$     &  $m_{l,3}^{\ast}$    & 1.7$m_0$    \\
 $\Delta_{3'}$  & 0.1    & eV     &  \,        & \,   & \,               &  $m_{t,3'}^{\ast}$   & -0.16$m_0$    \\
 \,             & \,     & \,     &  \,        & \,   & \,               &  $m_{l,3'}^{\ast}$   & 1.9$m_0$    \\ \hline \hline
\end{tabular}
\end{table}

\section{Wavevector Order Analysis of Intravalley Spin-Flip Transitions}
The theory for intravalley spin flips in Ge and Si share similar features. In Ref.~[\onlinecite{Song_arXiv12}], we have analyzed the case of Si. Here we summarize the important findings and discuss the difference for the case of Ge. By invoking space inversion and time reversal symmetries the leading order terms of intravalley scattering between $|\mathbf{k}_1=\mathbf{K}+\mathbf{q}/2,\,\Uparrow \rangle$ and $|\mathbf{k}_2=\mathbf{K}-\mathbf{q}/2,\,\Downarrow \rangle$ are found to be
\begin{eqnarray}
&& \!\!\!\! \frac{\mathbf{q}^{\otimes 2}}{8}\Big\langle \mathbf{K},\Downarrow \!\! \Big| (\bm{\mathcal{L}}^{\dagger})^{\otimes 2}\!\!\mathcal{A}^{+,\lambda}_{\mathbf{q}} +  \mathcal{A}^{+,\lambda}_{\mathbf{q}} \bm{\mathcal{L}}^{\otimes 2} - 2\bm{\mathcal{L}}^{\dagger} \mathcal{A}^{+,\lambda}_{\mathbf{q}} \bm{\mathcal{L}} \Big| \mathbf{K},\Uparrow \Big\rangle  \nonumber \\ && + \frac{\mathbf{q}}{2}\cdot \Big\langle \mathbf{K},\Downarrow \!\! \Big| \bm{\mathcal{L}}^{\dagger}\!\!\mathcal{A}^{-,\lambda}_{\mathbf{q}} +  \mathcal{A}^{-,\lambda}_{\mathbf{q}} \bm{\mathcal{L}} \Big| \mathbf{K}\Uparrow \Big\rangle\,. \label{eq:intra_order}
\end{eqnarray}
Lower order terms (in $\mathbf{q}$) vanish due to the celebrated Elliott-Yafet cancelation.\cite{Yafet_1963,Song_arXiv12} In this notation, $\mathbf{q}^{\otimes 2}\!\cdot\!\bm{\mathcal{L}}^{\otimes 2}$ denotes the scalar product of two second-rank tensors (each formed by a dyadic product of the vector with itself). $\bm{\mathcal{L}}$ is the derivative in $\mathbf{k}$-space with its components defined by
\begin{eqnarray}
\mathcal{L}_i | \mathbf{k},s \rangle  \equiv  \underset{\delta \mathbf{k} \rightarrow 0}{\text{lim}} \frac{| \mathbf{k}+\delta k_i,s \rangle - | \mathbf{k},s \rangle}{\delta k_i} \;. \label{eq:Lop}
\end{eqnarray}
In connection with the $L$~point Hamiltonian, $\bm{\mathcal{L}}$ operates on the eigenvectors [$\mathbf{C}_\gamma(\mathbf{k},s)$] and the envelope phase of the wavefunction [$\exp{(i\mathbf{k}\cdot\mathbf{r})}$]. The electron-phonon interaction in Eq.~(\ref{eq:intra_order}) is given by
\begin{eqnarray}
\mathcal{A}^{\pm,\lambda}_{\mathbf{q}}  = \bm{\xi}^{\pm,\lambda}_{\mathbf{q}} \cdot \bm{\nabla}\!\mathcal{V}_{\pm}\,\,, \label{eq:ei}
\end{eqnarray}
where the $+$ and $-$ signs denote, respectively, the in-phase and out-of-phase motion of atoms in the unit-cell. For scattering with long-wavelength acoustic phonon modes ($\lambda$ is TA or LA), the out-of-phase polarization vector ($\bm{\xi}^{-,\lambda}_{\mathbf{q}}$) is linear in $\mathbf{q}$ while the in-phase vector ($\bm{\xi}^{+,\lambda}_{\mathbf{q}}$) has a zeroth-order dependence (e.g., $q_i/q$ terms). It is the opposite case for scattering with long-wavelength optical phonon modes ($\lambda$ is TO or LO). These wavevector dependencies are taken into account in finding the power-law order of the intravalley spin-flip matrix element. Denoting the atoms positions in the unit cell by $\bm\tau_A$ and $\bm\tau_B$ with respect to the cell's origin, the potential in Eq.~(\ref{eq:ei}) reads
\begin{eqnarray}
\mathcal{V}_{\pm}(\mathbf{r}) \!\!\! & =& \!\!\! \mathcal{V}_{\rm{at}}(\mathbf{r}-\bm\tau_A) \pm \mathcal{V}_{\rm{at}}(\mathbf{r}-\bm\tau_B),\nonumber \\
\mathcal{V}_{\rm{at}}(\mathbf{r}) \!\!\! & =& \!\!\! V_{\rm{at}}(\mathbf{r}) \mathcal{I} + \frac{\hbar}{4m^2_0 c^2}\left[\bm{\nabla}V_{\rm{at}}(\mathbf{r})\!\times\!\mathbf{p} \right]\cdot \bm\sigma\,, \label{eq:potential}
\end{eqnarray}
where the spin-orbit interaction is included in the atomic potential.

In the next step of the analysis, we expand the states around the valley center. The \textit{bra} and \textit{ket} states in  Eq.~(\ref{eq:intra_order}) are taken at the average of $\mathbf{k}_1=\mathbf{K}+\mathbf{q}/2$ and $\mathbf{k}_2=\mathbf{K}-\mathbf{q}/2$. We expand this averaged state around the valley center position ($\mathbf{K}_0$),
\begin{eqnarray}
| \mathbf{K},s \rangle  =  | \mathbf{K}_0,s \rangle + \mathbf{K}\cdot\bm{\mathcal{L}}| \mathbf{K}_0,s \rangle + \mathcal{O}(K^2)\,, \label{eq:center}
\end{eqnarray}
where $\mathbf{K}$ is measured with respect to $\mathbf{K}_0$. Substituting this expansion in Eq.~(\ref{eq:intra_order}), one can identify which terms vanish. This identification is carried straightforwardly using the transformation properties of $\bm{\mathcal{L}}$, $\bm{\nabla}\!\mathcal{V}_{\pm}$ and $| \mathbf{K}_0,s \rangle$ under space inversion and time reversal symmetries. 

The difference between the analysis of Si and Ge stems from the position of the valley center. The valley center in Ge is at the zone-edge $L$~point, and in Si it is inside the Brillouin zone ($0.15 \times 2\pi/a$ away from the $X$~point along the $\Delta$-axis). Since $\mathbf{K}_0$ and $-\mathbf{K}_0$ are the same point in Ge, space inversion operation keeps $| \mathbf{K}_0,s \rangle$ invariant in Ge but not in Si. Together with the transformations of $\bm{\mathcal{L}}$, $\bm{\nabla}\!\mathcal{V}_{\pm}$, and $| \mathbf{K}_0,s \rangle$ one can readily identify the dominant contributions to intravalley spin-flip matrix elements. Scattering with long-wavelength acoustic phonons is led by $q_{\ell}q_m$ products in Si and by $K_{\ell}q_mq_n$ products in Ge. For scattering with long-wavelength optical phonons, the leading terms in Si are linear in $\mathbf{q}$, and in Ge they include $K_{\ell}q_m$ products. Finding the exact products, their coefficients and deformation potential constants requires a combination of $\mathbf{k}$$\cdot$$\mathbf{p}$, rigid-ion and group theories.\cite{Song_arXiv12}


\begin{thebibliography}{99}
\bibitem{footnote_gamma_j2} Terms that scale with $\gamma_j^2\mathcal{O}(k^2)$ were omitted since $\gamma_j \ll 1$.
\bibitem{footnote_optical} In Si, scattering with long-wavelength optical and acoustic phonons becomes comparable at room temperature (when the population of optical phonons is no longer negligible). The reason for the negligible contribution of optical modes to intravalley spin relaxation in Ge stems from the different symmetries of the electronic wavefunctions at the vicinity of the $L$~point (Ge) and the $\Delta$~axis (Si).
\bibitem{footnote_scattering_symmetries} Scattering symmetries that are inherent to the electronic states are kept. However, approximating the electron-phonon interaction by $\Xi q$ masks other scattering symmetries along certain high-symmetry directions. Replacing the scalar interaction with the correct tensorial form restores these scattering symmetries (see Ref.~[\onlinecite{Song_arXiv12}] for further details).
\end{thebibliography}
\end{document}